\documentclass[a4paper,onecolumn,unpublished]{quantumarticle}
\pdfoutput=1
\usepackage[pagebackref,colorlinks,hypertexnames=false]{hyperref}
\usepackage[utf8]{inputenc}
\usepackage[english]{babel}
\usepackage[T1]{fontenc}
\usepackage{amsmath}
\usepackage{bm}
\usepackage{amsthm}
\usepackage{amssymb}
\usepackage{graphicx}
\usepackage{booktabs}
\usepackage{float}
\usepackage{comment}

\usepackage[commandnameprefix=always]{changes}

\usepackage[numbers,sort&compress]{natbib}

\usepackage{tikz}
\usepackage{quantikz}

\tikzset{
        clifford/.style={very thick,draw=green!65!black,fill=green!15},
        noise/.style={very thick,rounded corners=4pt,draw=orange!80!black,fill=orange!15},
        nonclifford/.style={very thick,draw=red!65!black,fill=red!15},
        measurement/.style={very thick,draw=blue!70!black,fill=blue!15},
        overview flow/.style={-{Latex[length=2.6mm,width=1.8mm]},very thick,draw=black!65},
        overview item/.style={draw=black!40,fill=white,inner sep=4pt,align=center},
        active coord/.style={draw=red!65!black,fill=red!8,inner sep=4pt,align=center},
        dormant coord/.style={draw=green!55!black,fill=green!8,inner sep=4pt,align=center}
    }

\usetikzlibrary{arrows.meta,positioning,calc,fit}
\usepackage{adjustbox}

\usepackage{zref-clever}
\newcommand{\cref}[1]{\zcref{#1}}
\zcsetup{
    cap,
    nameinlink=false
}

\newcommand{\addtheorem}[1]{%
  \AddToHook{env/#1/begin}{%
    \zcsetup{countertype={theorem=#1}}%
  }%
  \zcRefTypeSetup{#1}{Name-sg={\MakeUppercase #1}}%
  \newtheorem{#1}[theorem]{\MakeUppercase #1}%
}

\addtheorem{axiom}
\addtheorem{lemma}
\addtheorem{corollary}
\addtheorem{proposition}
\addtheorem{fact}
\addtheorem{remark}

\usepackage{mathtools}

\DeclarePairedDelimiter\rbra{\lparen}{\rparen}
\DeclarePairedDelimiter\sbra{\lbrack}{\rbrack}
\DeclarePairedDelimiter\cbra{\{}{\}}
\DeclarePairedDelimiter\abs{\lvert}{\rvert}
\DeclarePairedDelimiter\Abs{\lVert}{\rVert}
\newcommand{\ketbra}[3][]{\left\lvert #2 \vphantom{#3} \right>_{#1}\!\!\left< #3 \vphantom{#2} \right\rvert}

\DeclareMathOperator{\tr}{tr}
\DeclareMathOperator{\wt}{wt}

\newcommand{\tFont}[1]{\texttt{#1}}
\newcommand{\symft}{\tFont{SymFT}}
\newcommand{\soft}{\tFont{SOFT}}
\newcommand{\stim}{\tFont{Stim}}
\newcommand{\tsim}{\tFont{Tsim}}
\newcommand{\clifft}{\tFont{Clifft}}
\newcommand{\syqma}{\tFont{SyQMA}}
\newcommand{\symphase}{\tFont{SymPhase.jl}}

\newcommand{\iFont}[1]{\textsf{#1}}
\newcommand{\iActiveDiagRot}{\iFont{ActiveDiagRot}}
\newcommand{\iActivePairRot}{\iFont{ActivePairRot}}
\newcommand{\iPromoteRot}{\iFont{PromoteRot}}
\newcommand{\iRandomMeas}{\iFont{RandomMeas}}
\newcommand{\iDetMeas}{\iFont{DetMeas}}
\newcommand{\iActiveDiagMeas}{\iFont{ActiveDiagMeas}}
\newcommand{\iActivePairMeas}{\iFont{ActivePairMeas}}

\newcommand{\cnot}{\ensuremath{\mathit{CNOT}}}
\newcommand{\cz}{\ensuremath{\mathit{CZ}}}
\renewcommand{\swap}{\ensuremath{\mathit{SWAP}}}

\newcommand{\cC}{\mathcal{C}}
\newcommand{\cD}{\mathcal{D}}
\newcommand{\cN}{\mathcal{N}}
\newcommand{\cO}{\mathcal{O}}
\newcommand{\cP}{\mathcal{P}}

\newcommand{\CC}{\mathbb{C}}
\newcommand{\FF}{\mathbb{F}}

\makeatletter
\newcommand{\myfirstpagefootnote}[1]{%
  \insert\footins{\footnotesize #1\par}%
}
\apptocmd{\@printauthors}{%
  \myfirstpagefootnote{\symft{} is the second-generation successor to \soft{}, see GitHub repository: \url{https://github.com/haoliri0/SOFT}.}}{}{}
\makeatother

\begin{document}

\title{SymFT: Universal Fault-Tolerant Quantum Circuit Simulation via Symbolic Clifford--Pauli Frames and Stabilizer Coordinates}

\author[1]{Wang Fang}
\email{njuwfang@gmail.com}
\author[2]{Huazhe Lou}
\author[3]{Riling Li}
\email{liriling@ict.ac.cn} \email{haoliri0@gmail.com}
\affil[1]{School of Informatics, University of Edinburgh}
\affil[2]{Arclight Quantum Computing Inc.}
\affil[3]{Institute of Computing Technology, Chinese Academy of Sciences}

\maketitle

\begin{abstract}
Fault-tolerant protocols often consist largely of stabilizer subcircuits, yet the non-Clifford operations required for universality make exact sampling costly.
We present \symft{}, a high-throughput simulator for Clifford-dominated circuits with Pauli rotations, stochastic Pauli noise, mid-circuit Pauli measurements, and measurement-record-controlled Pauli feedback.
It combines two ideas.
First, symbolic Clifford--Pauli frame factorization reduces branch-probability sampling to Pauli rotations and measurement projectors, with noise and feedback represented by symbolic signs.
Since the residual Clifford and Pauli frames are unitary, they do not affect branch probabilities and need not be applied in every shot.
Second, adaptive stabilizer-coordinate planning uses a shared stabilizer--destabilizer tableau to define the basis and stores only the active non-stabilizer degrees of freedom in a dynamically sized dense active-state vector.
It resolves basis changes once and emits direct multi-coordinate sampling instructions, thereby avoiding per-shot tableau updates and localization-induced Clifford transformations of the dense vector.
Across the tested pure-Clifford and near-Clifford circuits, \symft{} achieves state-of-the-art sampling performance.
On a single CPU core, it achieves a $2.51\text{--}2.56\times$ speedup over \stim{} for surface-code circuits and a $1.86\text{--}3.51\times$ speedup over \clifft{} for magic-state cultivation and distillation circuits.
For the tested cultivation circuits, its GPU sampling throughput also exceeds that of our previous simulator, \soft{}, by more than two orders of magnitude.
\end{abstract}

\section{Introduction}
\label{sec:intro}

Fault-tolerant quantum computation protects logical information by encoding it in quantum error-correcting codes and implementing logical operations so that a sufficiently small set of physical faults does not cause an uncorrectable logical error~\cite{Shor96,Gottesman97}.
Classical simulation is used throughout the design and validation of such fault-tolerant protocols, including verifying logical gadgets, estimating acceptance and logical-failure probabilities, and testing decoders and postselection strategies.

Classical simulation becomes more challenging when fault-tolerant protocols incorporate the logical non-Clifford operations needed for universality.
Clifford gates and Pauli measurements admit efficient simulation within the stabilizer formalism~\cite{Gottesman98}, but alone are not computationally universal.
Fault-tolerant protocols realize these non-Clifford operations through additional constructions, including magic-state injection and distillation~\cite{BK05,GF19,Litinski19,HIF24,WHY24,LTF+25,ITH+25}; code switching or gauge fixing to access transversal non-Clifford gates~\cite{PR13,ADP14,Hector15,KB15}; concatenated-code schemes~\cite{JL14,CJL16}; pieceably fault-tolerant logical gates~\cite{TYC16,TYC17}; and, more recently, magic-state cultivation~\cite{GSJ24,HTI+25,CCL+26,VJG+26,STC+26,HBW26,CFS26}.

Although these approaches implement non-Clifford operations differently, the resulting fault-tolerant circuits often share a common structure from the perspective of classical simulation.
They are typically dominated by Clifford gates and Pauli measurements, with a smaller number of essential non-Clifford operations.
They may also include circuit-level noise, measurement-record-controlled Pauli feedback, and detectors or postselection conditions defined by measurement outcomes.
Exact simulation must exploit the stabilizer structure while retaining the coherent amplitudes created by non-Clifford operations.
The cost of representing the remaining non-stabilizer part in sampling depends on the chosen method and may be governed by the number of non-Clifford gates, the stabilizer rank, the number of nonzero generalized-stabilizer coefficients, or the dimension of a dense active-state vector.

Stabilizer simulators such as \stim{} reuse Clifford preprocessing across shots~\cite{Gidney21}.
Universal circuits additionally require a representation of the non-Clifford state.
Stabilizer-rank methods expand a non-stabilizer resource as a sum of stabilizer states~\cite{BG16,BBC+19,QPG21}, while ZX-calculus methods, including \tsim{}, simplify the global diagram and then seek a small sum-over-Clifford expansion~\cite{KW22,KWV22,SK24,SK25,HLZ26}.
In the generalized-stabilizer representation used by \soft{}~\cite{LZZ+25}, a stabilizer--destabilizer tableau defines an orthonormal basis, and an arbitrary state is represented by coefficients in that basis~\cite{Yoder12}.
\clifft{} instead confines the non-stabilizer degrees of freedom to a dynamically sized dense active-state vector over $k$ active virtual qubits~\cite{CL26}.
These approaches exploit different structures: global ZX simplification can reduce a stabilizer decomposition, sparse generalized-stabilizer storage can exploit low coefficient support, and a dense active-state representation can exploit the creation and removal of non-stabilizer degrees of freedom.

\subsection{SymFT overview}
Against this background, we present \symft{}, a high-throughput simulator for the noisy, adaptive Clifford-dominated circuits considered here.
\symft{} is the second-generation successor to our earlier simulator \soft{}~\cite{LZZ+25}.
It replaces \soft{}'s per-shot sparse generalized-stabilizer evolution with symbolic Clifford--Pauli frame factorization shared across shots and adaptive planning over stabilizer coordinates, compiling direct kernels for a dynamically sized dense active-state vector.

\paragraph{Symbolic Clifford--Pauli frame factorization.}
For an $n$-qubit circuit initialized in $\ket{0^n}$, let $\bm{s}$ denote the Pauli noise choices, $\bm{m}$ the measurement record, and $K\rbra{\bm{s},\bm{m}}$ the corresponding measurement branch operator.
The first stage of \symft{} factorizes this operator as
\begin{equation*}
    K\rbra{\bm{s},\bm{m}}
    \doteq
    C E\rbra{\bm{s},\bm{m}}\,
    \cO\rbra{\bm{s},\bm{m}},
\end{equation*}
where $\doteq$ denotes equality up to global phase, $C$ is a concrete Clifford frame, $E\rbra{\bm{s},\bm{m}}$ is a symbolic Pauli frame, and $\cO\rbra{\bm{s},\bm{m}}$ is the ordered sequence of Pauli rotations and Pauli measurement projectors pulled back through the two frames in the Heisenberg picture.
Because the residual factor $C E\rbra{\bm{s},\bm{m}}$ is unitary for every assignment, it does not change the probability of the measurement branch:
\begin{equation*}
    \Pr\rbra{\bm{m}\mid\bm{s}}
    = \Abs*{\cO\rbra{\bm{s},\bm{m}} \ket{0^n}}^2.
\end{equation*}
Thus, when sampling branch probabilities, \symft{} need not revisit Clifford gates, Pauli noise operations, or Pauli feedback controlled by the measurement record.
The effect of the Clifford gates is already contained in the pulled-back Pauli operators, whereas Pauli noise and feedback remain only as symbolic signs on the rotations and projectors.
\emph{This construction is related to symbolic methods~\cite{BSH+21,FY24a,FY24b,UA26}, but confines shot-dependent symbolic data to these signs}.

\paragraph{Adaptive stabilizer-coordinate planning.}
The second stage of \symft{} is a planning pass that compiles the residual evolution $\cO\rbra{\bm{s},\bm{m}}\ket{0^n}$ into an ordered sequence of sampling instructions.
During planning, \symft{} maintains a canonical stabilizer--destabilizer tableau
\begin{equation*}
    \overline{Z}_1,\ldots,\overline{Z}_n,
    \qquad
    \overline{X}_1,\ldots,\overline{X}_n.
\end{equation*}
The stabilizer--destabilizer tableau defines a stabilizer coordinate system in which each generator pair specifies one coordinate, or equivalently one virtual qubit.
The first $k$ stabilizer coordinates are active, and the remaining $n-k$ are dormant.
Let $\ket{\overline{0}}$ denote the joint $+1$ eigenstate of the current stabilizer generators.
The state represented during sampling then has the form
\begin{equation*}
    \ket{\psi} = \sum_{x\in\FF_2^k}
    \alpha_x \overline{X}_1^{x_1}\cdots \overline{X}_k^{x_k} \ket{\overline 0} \leftrightarrow
    \ket{\alpha}_A\in\CC^{2^k}.
\end{equation*}
Thus, a single tableau on $n$ physical qubits defines the shared stabilizer basis, while a dense active-state vector of dimension $2^k$ stores the coefficients of the active-basis states.
\emph{Compared with per-shot sparse generalized-stabilizer implementations such as \soft{}~\cite{LZZ+25}, \symft{} stores this vector contiguously but computes the tableau trajectory and basis changes only once during planning}.

For each pulled-back rotation or measurement, the planner decomposes its Pauli generator or observable into active and dormant components in this coordinate system.
Dormant components are handled through tableau updates and symbolic Pauli corrections, whereas active components are compiled into specialized instructions on the dense active-state vector; any required basis changes are resolved during planning.
A non-Clifford rotation can promote one dormant stabilizer coordinate, while an active measurement can return one stabilizer coordinate to the dormant set.
\emph{Unlike \clifft{}, which uses Clifford localization to map an active Pauli to a single virtual axis~\cite{CL26}, \symft{} emits a direct diagonal or paired-amplitude update for a multi-coordinate Pauli.
This avoids localization-induced Clifford transformations of the dense active-state vector during sampling}.

After planning, the \symft{} sampler executes only the resulting instructions: it samples Pauli noise choices, evaluates symbolic signs, updates the dense active-state vector, and records measurement outcomes.
In particular, it does not revisit Clifford gates, update Pauli frames, conjugate Pauli strings, or reconstruct the planning tableau.

\subsection{Contributions and organization}
The main contributions of this work are as follows:
\begin{itemize}
    \item We introduce an exact symbolic Clifford--Pauli frame factorization that removes residual unitary frames from branch-probability sampling.
    \item We introduce adaptive stabilizer-coordinate planning, which combines a shared tableau with a dynamically sized dense active-state vector and emits direct multi-coordinate sampling instructions.
    \item We provide \symft{} as a Python package with C\texttt{++} CPU and CUDA GPU backends.
    Benchmarks show higher single-core CPU throughput than \stim{} on pure-Clifford QEC and than \clifft{} and \tsim{} on magic-state cultivation, as well as higher GPU throughput than \tsim{} and \soft{} on cultivation.
\end{itemize}

The remainder of the paper is organized as follows.
\cref{sec:circuit-model} defines the noisy circuit model.
\cref{sec:factorization} develops the symbolic Clifford--Pauli frame factorization.
\cref{sec:planning} presents adaptive stabilizer-coordinate planning.
\cref{sec:sampling} gives the sampling algorithm and its complexity.
\cref{sec:implementation} describes the implementation.
\cref{sec:benchmark} reports the benchmark results.
\cref{sec:related-work} discusses related work, and \cref{sec:conclusion} concludes with future directions.

\section{Noisy Circuit Model and Conventions}
\label{sec:circuit-model}

We use the Pauli group as a common algebraic language for gates, measurements, and noise.
Let $\cP_n$ denote the group of $n$-qubit Pauli operators.
Each $P \in \cP_n$ is a tensor product of operators from $\{I,X,Y,Z\}$, multiplied by a phase in $\{\pm 1,\pm i\}$.
Whenever a Pauli operator is used as a rotation generator, measurement observable, or stochastic noise operator, we restrict it to be Hermitian, equivalently $P^2=I$.

We consider noisy quantum circuits in which Clifford operations constitute most of the circuit, while non-Clifford operations, measurements, and noise admit Pauli-based representations.
More precisely, \symft{} supports the following families of operations:
\begin{itemize}
    \item \textbf{Clifford gates.}
    The $n$-qubit Clifford group $\cC_n = \cbra{C  \mid \forall P \in \cP_n, CPC^{\dagger} \in \cP_n}$ consists of unitaries that normalize the Pauli group under conjugation. The group is generated by $H$, $S$, and $\cnot{}$, and includes all single-qubit Clifford gates as well as two-qubit gates such as $\cz{}$ and $\swap{}$.
    \item \textbf{Exact non-Clifford Pauli rotations.}
    A rotation generated by a Hermitian multi-qubit Pauli operator is written as
    \begin{equation*}
        R_P\rbra{\theta} = \exp\rbra*{-i\theta P/2} = \cos\rbra{\theta/2} I - i\sin\rbra{\theta/2} P, \qquad P \in \cP_n,\quad P^2=I.
    \end{equation*}
    This family includes $T$ and $T^\dagger$ gates, up to global phase, as the cases $P=Z$ and $\theta=\pm\pi/4$.
    \item \textbf{Projective Pauli measurements.}
    A measurement of a Hermitian multi-qubit Pauli observable $P$ has projectors
    \begin{equation*}
        \Pi_{P,0} = \frac{I+P}{2}, \quad \Pi_{P,1} = \frac{I-P}{2}, \qquad P \in \cP_n,\quad P^2=I.
    \end{equation*}
    The binary outcome may be recorded and used by later operations, allowing both terminal and mid-circuit measurements.
    \item \textbf{Stochastic Pauli noise.}
    A stochastic multi-qubit Pauli channel acts as
    \begin{equation*}
        \cN\rbra{\rho} = \sum_j p_j N_j \rho N_j, \qquad p_j \geq 0,\quad \sum_j p_j = 1, \quad N_j \in \cP_n, \quad N_j^2=I.
    \end{equation*}
    This family includes independent and correlated Pauli errors, as well as multi-qubit depolarizing noise.
    \item \textbf{Measurement-record-controlled Pauli feedback.}
    A multi-qubit Pauli operator may be applied conditionally on the parity of a specified subset of previously recorded measurement outcomes.
    % \item \textbf{Detector and observable annotations.}
    % Detectors and logical observables may be defined as parities of recorded measurement outcomes. These annotations support syndrome extraction, postselection, and the evaluation of logical measurement results.
\end{itemize}
The model can be extended to more general classically controlled operations, including controls defined by Boolean functions other than parity and feedback implemented by non-Pauli gates.
Such extensions are generally less efficient: different measurement-record branches may require distinct quantum evolutions, and non-Pauli feedback introduces additional non-Clifford operations.

\section{Symbolic Clifford--Pauli Frame Factorization}
\label{sec:factorization}

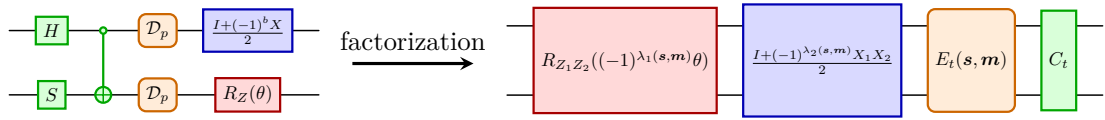
\begin{figure}[htb]
    \centering
    $\begin{gathered}
            \scalebox{.7}{
                \begin{quantikz}
                    & \gate[style={clifford}]{H} & \ctrl[style={clifford}]{1} & \gate[style=noise]{\cD_p} & \gate[style={measurement}]{\frac{I+ \rbra{-1}^b X}{2}} & \\
                    & \gate[style={clifford}]{S} & \targ[style={clifford}]{} & \gate[style=noise]{\cD_p} & \gate[style=nonclifford]{R_{Z}\rbra{\theta}} &
                \end{quantikz}
            }  \tikz{\draw[-stealth,very thick] (0,0) -- (1.6,0);\node at (.8,.3) {factorization};}
            \scalebox{.7}{
                \begin{quantikz}
                    & \gate[2,style=nonclifford]{R_{Z_1Z_2}\rbra{\rbra{-1}^{\lambda_1\rbra{\bm{s},\bm{m}}}\theta}} & \gate[2,style={measurement}]{\frac{I+ \rbra{-1}^{\lambda_2\rbra{\bm{s},\bm{m}}} X_1X_2}{2}} & \gate[2,style=noise]{E_t\rbra{\bm{s},\bm{m}}} & \gate[2,style=clifford]{C_t} & \\
                    & & & & &
                \end{quantikz}
            }
    \end{gathered}$
    \caption{Illustration of symbolic Clifford--Pauli frame factorization. The original circuit (left) contains Clifford gates, single-qubit depolarizing noise channels, a Pauli rotation, and a Pauli measurement. In the factorized circuit (right), the rotation and measurement have been pulled back to the beginning of the circuit. Clifford conjugation transforms their Pauli operators, while conjugation by the symbolic Pauli frame contributes the signs encoded by $\lambda_1\rbra{\bm{s},\bm{m}}$ and $\lambda_2\rbra{\bm{s},\bm{m}}$. The accumulated Pauli and Clifford operations remain in the residual frames $E_t\rbra{\bm{s},\bm{m}}$ and $C_t$, respectively, and the relative order of the pulled-back operations is preserved.}
    \label{fig:factorization-example}
\end{figure}

\symft{} begins with symbolic Clifford--Pauli frame factorization.
It traverses the circuit once in execution order.
During this traversal, Clifford operations accumulate in a Clifford frame, while stochastic Pauli noise and measurement-record-controlled Pauli feedback accumulate in a symbolic Pauli frame.
Each Pauli rotation and Pauli measurement is pulled back through the current frames and appended to an ordered sequence acting directly on the input state.
The relative order of these rotations and measurements is preserved.
At the end of this pass, the circuit has been factorized into the pulled-back operation sequence and the residual Clifford and symbolic Pauli frames, as illustrated in \cref{fig:factorization-example}.

The factorization relies on two elementary properties.
First, Clifford conjugation maps Pauli operators to Pauli operators: $C^\dagger P C \in \cP_n$ for $C \in \cC_n, P \in \cP_n$.
Second, conjugation by a Pauli operator can only change the sign of another Pauli operator.
Defining the anticommutation indicator
\begin{equation*}
    \omega(A,P) =
    \begin{cases}
        0, & AP=PA,\\
        1, & AP=-PA,
    \end{cases}
    \qquad A,P\in\cP_n,
\end{equation*}
we then have
\begin{equation*}
    A^\dagger P A = \rbra{-1}^{\omega(A,P)}P.
\end{equation*}
Consequently, pulling a Pauli rotation or measurement through a Clifford frame changes its Pauli operator, whereas pulling it through a symbolic Pauli frame changes only its sign, which is represented symbolically.
For example, consider a stochastic Pauli channel
\begin{equation*}
    \cN_t\rbra{\rho}
    =
    \sum_{j\in J_t} p_{t,j} N_{t,j}\rho N_{t,j}.
\end{equation*}
We introduce a symbolic noise variable $s_t\in J_t$ satisfying $\Pr\rbra{s_t=j}=p_{t,j}$
and write $N_t\rbra{s_t}=N_{t,s_t}$ for the selected Pauli noise operator.
Equivalently, one may use one-hot variables $s_{t,j}\in\{0,1\}$ satisfying $\sum_j s_{t,j}=1$.
In that representation,
\begin{equation}\label{eq:symbolic-noise-sign}
    N_t\rbra{\bm{s}_t}^{\dagger} P N_t\rbra{\bm{s}_t} =
    \rbra{-1}^{\bigoplus_{j\in J_t} s_{t,j}\omega\rbra{N_{t,j},P}} P.
\end{equation}
Thus, the effect of stochastic Pauli noise on a pulled-back Pauli operator is captured entirely by a symbolic sign.

We now give the complete factorization procedure.
Let $o_1,o_2,\ldots,o_T$ denote the circuit operations in execution order.
During factorization, \symft{} maintains the factorization
\begin{equation}\label{eq:factorization-invariant}
    K_t \doteq C_t E_t\rbra{\bm{s},\bm{m}}\,
    \cO_t\rbra{\bm{s},\bm{m}},
\end{equation}
where $\doteq$ denotes equality up to global phase.
The three factors have the following meanings:
\begin{itemize}
    \item $C_t$ is a concrete Clifford frame containing the accumulated Clifford gates;
    \item $E_t\rbra{\bm{s},\bm{m}}$ is a symbolic Pauli frame containing the accumulated Pauli operators arising from stochastic Pauli noise and measurement-record-controlled Pauli feedback, with symbolic variables specifying whether each operator is applied;
    \item $\cO_t\rbra{\bm{s},\bm{m}}$ is the ordered product of the Pauli rotations and Pauli measurement projectors that have been pulled back through the two frames.
\end{itemize}
Here $\bm{s}$ denotes the symbolic variables associated with stochastic Pauli noise channels, whereas $\bm{m}$ denotes the symbolic Pauli measurement outcomes.

Initially,
\begin{equation*}
    C_0=I,
    \qquad
    E_0\rbra{\bm{s},\bm{m}}=I,
    \qquad
    \cO_0\rbra{\bm{s},\bm{m}}=I.
\end{equation*}
The circuit is then processed once in execution order.
For each operation $o_t$, the frames are updated according to the following rules, where only the components that change are displayed:
\begin{itemize}
    \item For a Clifford gate $g$,
    \begin{equation*}
        C_t = g C_{t-1}.
    \end{equation*}
    \item For a stochastic Pauli noise channel $\cN_t: \rho \mapsto \sum_{j\in J_t}p_{t,j}N_{t,j}\rho N_{t,j}$, the symbolic Pauli frame is updated as
    \begin{equation*}
        E_t\rbra{\bm{s}, \bm{m}} = \prod_{j\in J_t}\rbra*{C_{t-1}^{\dagger} N_{t,j} C_{t-1}}^{s_{t,j}} E_{t-1}\rbra{\bm{s},\bm{m}},
    \end{equation*}
    For each occurrence of the channel, the variables $\cbra{s_{t,j}:j\in J_t}$ are one-hot: exactly one equals $1$, with $\Pr\rbra{s_{t,j}=1}=p_{t,j}$, and all others equal $0$.
    \item For a Pauli rotation $R_{P}\rbra{\theta}$,
    \begin{equation*}
        \cO_{t}\rbra{\bm{s},\bm{m}} = R_{C^{\dagger}_{t-1}P C_{t-1}}\rbra*{\rbra{-1}^{\lambda_t\rbra{\bm{s},\bm{m}}}\theta}\cO_{t-1}\rbra{\bm{s},\bm{m}},
    \end{equation*}
    where the affine Boolean expression $\lambda_t\rbra{\bm{s},\bm{m}}$ is defined by
    \begin{equation*}
        E_{t-1}\rbra{\bm{s}, \bm{m}}^{\dagger} \rbra*{C^{\dagger}_{t-1}P C_{t-1}} E_{t-1}\rbra{\bm{s}, \bm{m}} = \rbra{-1}^{\lambda_t\rbra{\bm{s},\bm{m}}} \rbra*{C^{\dagger}_{t-1}P C_{t-1}}
    \end{equation*}
    as in \cref{eq:symbolic-noise-sign}.
    \item For a Pauli measurement with observable $P$, write the projector associated with symbolic outcome $m_t \in \cbra{0,1}$ as $\Pi_{P, m_t} = \frac{I+\rbra{-1}^{m_t}P}{2}$. Using the same definition of $\lambda_t$ as above, the pulled-back projector is
    \begin{equation*}
        \cO_{t}\rbra{\bm{s},\bm{m}} = \Pi_{C_{t-1}^{\dagger}P C_{t-1}, m_t\oplus \lambda_t\rbra{\bm{s},\bm{m}}} \cO_{t-1}\rbra{\bm{s},\bm{m}}.
    \end{equation*}
    \item For a measurement-record-controlled Pauli operator $P^{e_t\rbra*{\bm{m}_{<t}}}$, where $e_t\rbra*{\bm{m}_{<t}}$ is the parity of a specified subset of earlier symbolic measurement outcomes,
    \begin{equation*}
        E_{t}\rbra{\bm{s},\bm{m}} = \rbra{C_{t-1}^{\dagger}PC_{t-1}}^{e_{t}\rbra*{\bm{m}_{<t}}} E_{t-1}\rbra{\bm{s},\bm{m}}
    \end{equation*}
\end{itemize}
After the final operation, the factorization becomes
\begin{equation}\label{eq:factorized-circuit}
    K_T \doteq C_T\, E_T\rbra{\bm{s},\bm{m}}\, \cO_T\rbra{\bm{s},\bm{m}}.
\end{equation}
For an input state $\rho_0$, the probability of a measurement branch $\bm{m}$, conditioned on the sampled noise choices $\bm{s}$, is
\begin{equation}\label{eq:factorized-branch-probability}
    \Pr\rbra*{\bm{m} \mid \bm{s}} = \tr\rbra*{K_T \rho_0 K_T^\dagger} = \tr\rbra*{\cO_T\rbra{\bm{s},\bm{m}} \rho_0 \cO_T\rbra{\bm{s},\bm{m}}^{\dagger}}.
\end{equation}
The second equality holds because $C_TE_T\rbra{\bm{s},\bm{m}}$ is unitary for every concrete assignment of $\bm{s}$ and $\bm{m}$.
Consequently, the residual Clifford and Pauli frames need not be applied to the quantum state when computing measurement-record statistics.
For the default input
$\rho_0=\ketbra{0}{0}^{\otimes n}$, the simulator only needs to execute the ordered factorized sequence
$\cO_T\rbra{\bm{s},\bm{m}}$.

The resulting representation is Heisenberg-like: Clifford gates and Pauli noise operations accumulate toward the end of the circuit, while Pauli rotations and measurements are pulled back toward its beginning.
Since both Clifford and Pauli conjugation preserve the Pauli form, this factorization requires only a single pass over the circuit.
Once it is complete, the noise choices $\bm{s}$ can be sampled from their specified distributions and the measurement outcomes $\bm{m}$ generated sequentially from their conditional probabilities, without another traversal of the original Clifford gates or Pauli noise operations.

\section{Planning Symbolic Pauli Rotations and Measurements}
\label{sec:planning}

After Clifford--Pauli frame factorization, only the symbolic sequence $\cO_T\rbra{\bm{s},\bm{m}}$ remains to act on $\ket{0^n}$.
In its second stage, \symft{} compiles this sequence into sampling instructions.

This planning procedure builds on the generalized-stabilizer simulation used by \soft{}~\cite{LZZ+25} and the active-state representation of \clifft{}~\cite{CL26}, but it neither performs localization nor emits localization-induced Clifford operations on the dense active-state vector.
Instead, a stabilizer--destabilizer tableau defines the stabilizer coordinate system from which the planner derives direct instructions on this vector.
The affine Boolean expressions introduced during factorization carry the effects of noise, feedback, and earlier measurements into the symbolic signs of these instructions.

The planner maintains only three pieces of data:
\begin{itemize}
    \item a stabilizer--destabilizer tableau representing the active and dormant stabilizer coordinates;
    \item the active width $k$, equal to the number of active stabilizer coordinates;
    \item an ordered sequence of sampling instructions for the later sampler.
\end{itemize}
The planner neither stores nor updates a dense active-state vector, and the sampler never reconstructs the tableau.
It instead executes the emitted instructions and updates a dense active-state vector of dimension $2^k$ using the sampled noise choices and measurement record.

At any point in the planning pass, let the current tableau have stabilizer generators
\begin{equation*}
    \overline{Z}_1,\overline{Z}_2,\ldots,\overline{Z}_n \in \cP_n
\end{equation*}
and destabilizer generators
\begin{equation*}
    \overline{X}_1,\overline{X}_2,\ldots,\overline{X}_n \in \cP_n,
\end{equation*}
satisfying
\begin{equation*}
    \overline{Z}_i\overline{Z}_j = \overline{Z}_j\overline{Z}_i,
    \qquad
    \overline{X}_i\overline{X}_j = \overline{X}_j\overline{X}_i,
    \qquad
    \overline{Z}_i\overline{X}_j = \rbra{-1}^{\delta_{i,j}}\overline{X}_j\overline{Z}_i.
\end{equation*}
Each generator pair $(\overline{Z}_j,\overline{X}_j)$ defines one stabilizer coordinate, or equivalently one virtual qubit.
The first $k$ stabilizer coordinates are active, and the remaining $n-k$ are dormant.
Let $\ket{\overline{0}}$ be the joint $+1$ eigenstate of $\overline{Z}_1,\ldots,\overline{Z}_n$.
The active basis is defined by
\begin{equation}\label{eq:active-basis}
    \ket{\overline{x}}_A = \overline{X}_1^{x_1}\cdots
    \overline{X}_k^{x_k}
    \ket{\overline{0}},
    \qquad
    x\in\FF_2^k.
\end{equation}
In this basis,
\begin{equation}\label{eq:active-logical-z}
    \overline{Z}_j\ket{\overline{x}}_A = \rbra{-1}^{x_j}\ket{\overline{x}}_A, \qquad 1\leq j\leq k,
\end{equation}
while the dormant stabilizers $\overline{Z}_{k+1},\ldots,\overline{Z}_n$ have eigenvalue $+1$ on every active-basis state $\ket{\overline{x}}_A$.

During sampling, the represented state and its dense active-state vector satisfy
\begin{equation}\label{eq:active-vector-representation}
    \cO_t\rbra{\bm{s},\bm{m}}\ket{0^n} =
    \sum_{x\in\FF_2^k} \alpha_x\ket{\overline{x}}_A \leftrightarrow
    \ket{\alpha}_A\in\CC^{2^k}.
\end{equation}
During planning, this equation serves only as a representation invariant: the planner does not store or update the coefficients $\alpha_x$.
The coefficients specify the semantics of the emitted instructions and are updated only later, during sampling.
At the beginning,
\begin{equation*}
    \ket{0^n} = \ket{\overline{0}}, \quad \overline{Z}_j = Z_j, \quad \overline{X}_j = X_j, \quad k = 0,
\end{equation*}
so the dense active-state vector is the scalar $1$.

\paragraph{Active/dormant Pauli decomposition.}
For a Hermitian Pauli operator $P$, write its decomposition in the current stabilizer coordinates as
\begin{equation}\label{eq:active-dormant-decomposition}
    P = \mu P_AP_D
\end{equation}
with
\begin{align*}
    P_A &= \overline{X}_A^a\overline{Z}_A^b =
    \overline{X}_1^{a_1}\cdots\overline{X}_k^{a_k} \overline{Z}_1^{b_1}\cdots\overline{Z}_k^{b_k},
    \\
    P_D &= \overline{X}_D^d\overline{Z}_D^c = \overline{X}_{k+1}^{d_1}\cdots\overline{X}_{n}^{d_{n-k}} \overline{Z}_{k+1}^{c_1}\cdots\overline{Z}_{n}^{c_{n-k}},
\end{align*}
where
\begin{equation*}
    a,b\in\FF_2^k,
    \qquad
    c,d\in\FF_2^{n-k},
    \qquad
    \mu\in\cbra{\pm1,\pm i}
\end{equation*}
are determined by commutation with the tableau generators: $a_j=1$ iff $P$ anticommutes with $\overline{Z}_j$, and $b_j=1$ iff $P$ anticommutes with $\overline{X}_j$; the dormant bits $d,c$ are obtained in the same way.

We say the dormant component is diagonal in the dormant stabilizer basis exactly when $d=0$.
In that case, $P$ preserves the active subspace and acts on the active basis as
\begin{equation}\label{eq:active-logical-action}
    P\ket{\overline{x}}_A = \mu P_A \ket{\overline{x}}_A = \mu \rbra{-1}^{b\cdot x}\ket{\overline{x\oplus a}}_A
\end{equation}

\subsection{Planning Pauli Rotations}
Consider a factorized Pauli rotation $R_{P}\rbra*{\rbra{-1}^{\lambda\rbra{\bm{s},\bm{m}}}\theta}$.
During planning, the Pauli $P$ is decomposed in the current active/dormant stabilizer coordinates as $P = \mu \overline{X}^a_A\overline{Z}^b_A\overline{X}^d_D\overline{Z}^c_D $ as in \cref{eq:active-dormant-decomposition}.

\paragraph{Dormant-nondiagonal rotations.}
Suppose $d \neq 0$.
Then $P$ anticommutes with at least one dormant stabilizer and maps the current active subspace to an orthogonal dormant part.
The rotation therefore promotes one dormant stabilizer coordinate and increases the active width by one.

Choose a dormant pivot $\overline{Z}_{h}$, $h > k$, such that $P$ anticommutes with $\overline{Z}_{h}$, equivalently $d_{h-k} = 1$.
The planner then performs a symplectic coordinate change on the tableau using this dormant pivot.
Concretely, for every generator other than the pivot, update
\begin{align*}
    \overline{Z}_j'
    &=
    \overline{Z}_j
    \overline{Z}_{h}^{a_j},
    &
    \overline{X}_j'
    &=
    \overline{X}_j
    \overline{Z}_{h}^{b_j},
    &
    1\leq j \leq k, \\
    \overline{Z}_{k+j}'
    &=
    \overline{Z}_{k+j}
    \overline{Z}_{h}^{d_{j}},
    &
    \overline{X}_{k+j}'
    &=
    \overline{X}_{k+j}
    \overline{Z}_{h}^{c_j},
    &
    1\leq j \leq n-k, j \neq h-k,
\end{align*}
and set the pivot pair to
\begin{equation*}
    \overline{Z}_h' = \overline{Z}_{h}, \qquad \overline{X}_h' = P.
\end{equation*}
These updates make every non-pivot tableau generator commute with $P$, preserve the canonical stabilizer--destabilizer commutation relations, and leave the active basis unchanged because
\begin{equation*}
    \ket{\overline{x}}_A = \overline{X}_1^{x_1}\cdots
    \overline{X}_k^{x_k}
    \ket{\overline{0}} = \rbra*{\overline{X}_1\overline{Z}_{h}^{b_1}}^{x_1}\cdots
    \rbra*{\overline{X}_k \overline{Z}_{h}^{b_k} }^{x_k}
    \ket{\overline{0}},
\end{equation*}
where the second equality holds because $\overline{Z}_{h}$ has eigenvalue $+1$ on $\ket{\overline{0}}$.
Finally, if $h\neq k+1$, the planner swaps the $h$th and $(k+1)$st generator pairs.
Thus, in the new stabilizer coordinates,
\begin{equation*}
    P = \overline{X}_{k+1}.
\end{equation*}
Because this coordinate change uses a dormant pivot, the old active basis embeds into the new active basis as follows:
\begin{equation}\label{eq:rotation-promotion-basis}
    \ket{\overline{x}}_A^{\text{old}} = \ket{\overline{x0}}_A^{\text{new}}, \qquad P\ket{\overline{x}}_A^{\text{old}} = \ket{\overline{x1}}_A^{\text{new}}.
\end{equation}
These identities hold because the new $\overline{Z}_{k+1}$ is a dormant stabilizer and therefore has eigenvalue $+1$ on every old active-basis state.

The planner increments the active width $k$ by one and emits
\begin{equation*}
    \iPromoteRot\rbra*{\theta, \lambda\rbra{\bm{s}, \bm{m}}}.
\end{equation*}
The semantics of the emitted instruction is
\begin{equation*}
    \ket{\alpha}_A \gets \exp\rbra*{-i\rbra{-1}^{\lambda\rbra{\bm{s},\bm{m}}}\theta\, \overline{X}_{k+1}/2}\ket{\alpha}_A.
\end{equation*}
Using \cref{eq:rotation-promotion-basis}, the promoted state is
\begin{equation*}
    \ket{\alpha}_A \gets \sum_{x\in \FF_2^k} \alpha_x\cos\rbra{\theta/2}\ket{\overline{x0}}_A - i\rbra{-1}^{\lambda\rbra{\bm{s},\bm{m}}}\sin\rbra{\theta/2}\alpha_x\ket{\overline{x1}}_A.
\end{equation*}
Equivalently, during sampling the dense active-state vector is updated as follows:
\begin{equation*}
    \alpha_{x0}' = \cos\rbra{\theta/2}\alpha_x, \qquad \alpha_{x1}' = - i\rbra{-1}^{\lambda\rbra{\bm{s},\bm{m}}}\sin\rbra{\theta/2}\alpha_x.
\end{equation*}
Thus, a dormant-nondiagonal rotation increases the dimension of the dense active-state vector from $2^k$ to $2^{k+1}$.
The stabilizer coordinates are updated during planning, while the sampler only executes the emitted promotion instruction.

\paragraph{Dormant-diagonal rotations.}
Suppose $d = 0$ and $a=b=0$.
Then, $P$ acts as a scalar on the represented state.
Since $P$ is Hermitian, this scalar is $\pm 1$, and the rotation contributes only a global phase.
No sampling instruction is therefore required.

\paragraph{Active-diagonal rotations.}
Suppose $d = 0$, $a = 0$, and $b\neq 0$.
Then, $P$ changes only the phases of the active-basis states, as shown in \cref{eq:active-logical-action}.
The planner emits an active-diagonal rotation instruction
\begin{equation*}
    \iActiveDiagRot\rbra{b, \mu, \theta, \lambda\rbra{\bm{s},\bm{m}}}.
\end{equation*}
The semantics of the emitted instruction is
\begin{equation*}
    \ket{\alpha}_A \gets \exp\rbra*{-i\rbra{-1}^{\lambda\rbra{\bm{s},\bm{m}}}\theta\, \mu \overline{Z}^b_A/2}\ket{\alpha}_A.
\end{equation*}
Using \cref{eq:active-logical-action}, the update to the dense active-state vector during sampling is
\begin{equation*}
    \alpha_x \gets \exp\rbra*{-i\rbra{-1}^{\lambda\rbra{\bm{s},\bm{m}}}\theta\, \mu \rbra{-1}^{b\cdot x}/2} \alpha_x.
\end{equation*}

\paragraph{Active-nondiagonal rotations.}
Suppose $d = 0$, $a \neq 0$.
Then, $P$ preserves the current active subspace, and by \cref{eq:active-logical-action}, the rotation acts only on the dense active-state vector.
The planner emits an active-rotation instruction
    \begin{equation*}
        \iActivePairRot\rbra*{a, b, \mu, \theta, \lambda\rbra{\bm{s},\bm{m}}}.
    \end{equation*}
    The tableau is unchanged. The semantics of the emitted instruction is
    \begin{equation*}
        \ket{\alpha}_A \gets \exp\rbra*{-i\rbra{-1}^{\lambda\rbra{\bm{s},\bm{m}}}\theta\, \mu \overline{X}^a_A\overline{Z}^b_A/2}\ket{\alpha}_A
    \end{equation*}
    Using \cref{eq:active-logical-action}, the update to the dense active-state vector during sampling is
    \begin{equation*}
        \alpha_x \gets \cos\rbra{\theta/2}\alpha_x - i \rbra{-1}^{\lambda\rbra{\bm{s},\bm{m}}} \sin\rbra{\theta/2} \mu\rbra{-1}^{b \cdot \rbra{x\oplus a}}\alpha_{x\oplus a}, \quad x\in\FF_2^k.
    \end{equation*}
Since $a \neq 0$, the basis states split into pairs $x_1 = x_0 \oplus a$.
On each pair, the update is
\begin{align*}
    \alpha_{x_0}' &= \cos\rbra{\theta/2} \alpha_{x_0} - i\rbra{-1}^{\lambda\rbra{\bm{s},\bm{m}}}\sin\rbra{\theta/2}\mu\rbra{-1}^{b\cdot x_1} \alpha_{x_1}, \\
    \alpha_{x_1}' &= \cos\rbra{\theta/2}\alpha_{x_1} - i\rbra{-1}^{\lambda\rbra{\bm{s},\bm{m}}}\sin\rbra{\theta/2}\mu\rbra{-1}^{b\cdot x_0} \alpha_{x_0}. 
\end{align*}

\subsection{Planning Pauli Measurements}
Consider a factorized Pauli measurement projector
\begin{equation*}
    \Pi_{P,m_t\oplus \lambda\rbra{\bm{s},\bm{m}}} = \frac{I+ \rbra{-1}^{m_t\oplus \lambda\rbra{\bm{s},\bm{m}}}P}{2},
\end{equation*}
where $m_t$ is the recorded measurement bit and $\lambda\rbra{\bm{s},\bm{m}}$ is the affine Boolean expression defining the symbolic sign induced by earlier noise variables, feedback operations, and measurement outcomes.
We distinguish the recorded bit $m_t$ from the effective eigenspace label:
\begin{equation}
    y_t = m_t\oplus \lambda\rbra{\bm{s},\bm{m}}.
\end{equation}
With this notation, the projector becomes
\begin{equation*}
    \Pi_{P,y_t} = \frac{I+ \rbra{-1}^{y_t}P}{2}
\end{equation*}

During planning, the Pauli $P$ is decomposed in the current active/dormant stabilizer coordinates as $P = \mu \overline{X}^a_A\overline{Z}^b_A\overline{X}^d_D\overline{Z}^c_D $ as in \cref{eq:active-dormant-decomposition}.

\paragraph{Dormant-nondiagonal random measurements.}
Suppose $d \neq 0$.
Then, $P$ anticommutes with at least one dormant stabilizer.
As in the dormant-nondiagonal rotation case, choose a dormant pivot $\overline{Z}_{h}$, $h > k$, such that $P$ anticommutes with $\overline{Z}_{h}$, equivalently $d_{h-k} = 1$.
The planner then performs a symplectic coordinate change on the tableau using this dormant pivot.
Concretely, for every generator other than the pivot, update
\begin{align*}
    \overline{Z}_j'
    &=
    \overline{Z}_j
    \overline{Z}_{h}^{a_j},
    &
    \overline{X}_j'
    &=
    \overline{X}_j
    \overline{Z}_{h}^{b_j},
    &
    1\leq j \leq k, \\
    \overline{Z}_{k+j}'
    &=
    \overline{Z}_{k+j}
    \overline{Z}_{h}^{d_{j}},
    &
    \overline{X}_{k+j}'
    &=
    \overline{X}_{k+j}
    \overline{Z}_{h}^{c_j},
    &
    1\leq j \leq n-k, j \neq h-k
\end{align*}
and set the pivot pair to
\begin{equation*}
    \overline{Z}_h'= \overline{Z}_h, \qquad \overline{X}_h' = P.
\end{equation*}
These updates make every non-pivot tableau generator commute with $P$, preserve the canonical stabilizer--destabilizer commutation relations, and leave the active basis unchanged.
Then,
\begin{align*}
    \Pi_{P,y_t}\ket{\alpha}_A &= \frac{I+ \rbra{-1}^{y_t}\,\overline{X}_{h}'}{2}\ket{\alpha}_A \\
    &= \frac{I+ \rbra{-1}^{y_t}\,\overline{X}_{h}'}{2}\sum_{x \in {\FF_2^k}}{\alpha_x}  \rbra*{\overline{X}_1'}^{x_1}\cdots
    \rbra*{\overline{X}_k'}^{x_k}\ket{\overline{0}} \\
    &= \sum_{x \in {\FF_2^k}}{\alpha_x}  \rbra*{\overline{X}_1'}^{x_1}\cdots
    \rbra*{\overline{X}_k'}^{x_k} \cdot \frac{I+ \rbra{-1}^{y_t}\,\overline{X}_{h}'}{2}\ket{\overline{0}}. \tag{{by $\overline{X}_j' \overline{X}_{h}' = \overline{X}_h' \overline{X}_{j}'$}}
\end{align*}
Since $P = \overline{X}_{h}'$ anticommutes with the dormant stabilizer $\overline{Z}_{h}'$, we have
\begin{equation*}
    \bra{\alpha}_{A} \overline{X}_{h}' \ket{\alpha}_A = 0,
\end{equation*}
and therefore both effective outcomes have probability $1/2$.
For any fixed assignment of previous symbols, the map $m_t \mapsto y_t = m_t\oplus \lambda\rbra{\bm{s},\bm{m}}$ is a bit flip.
Hence, the recorded bit $m_t$ itself is uniformly random:
\begin{equation*}
    \Pr\rbra{m_t} = \frac{1}{2}.
\end{equation*}
The planner emits a random-measurement instruction
\begin{equation*}
    \iRandomMeas\rbra{m_t},
\end{equation*}
which samples and records $m_t$ with equal probability during sampling.

To restore canonical post-measurement form, the pivot stabilizer and destabilizer are exchanged:
\begin{equation*}
    \overline{Z}_{h}'' = \overline{X}_{h}' = P, \qquad \overline{X}_{h}'' = \overline{Z}_{h}' = \overline{Z}_h.
\end{equation*}
The updated tableau therefore represents the $+1$ eigenspace of $P$.
The branch with effective outcome $y_t$ is represented by the symbolic Pauli correction
\begin{equation*}
    \rbra*{\overline{X}_h''}^{y_t} = \rbra*{\overline{X}_h''}^{m_t\oplus \lambda\rbra{\bm{s},\bm{m}}},
\end{equation*}
which is propagated into the remaining Pauli rotations and measurements by updating their symbolic signs.
The dense active-state vector is unchanged, as is the active width $k$.

\paragraph{Dormant-diagonal deterministic measurements.}
Suppose $d = 0$ and $a = b = 0$.
Then, $P$ acts as a scalar on the represented state, i.e.,
\begin{equation*}
    P\ket{\alpha}_A = \mu \ket{\alpha}_A.
\end{equation*}
Since $P$ is Hermitian, write this scalar as
\begin{equation*}
    \mu=\rbra{-1}^{\eta}, \qquad \eta\in\FF_2.
\end{equation*}
Then,
\begin{equation*}
    \Pi_{P,m_t\oplus \lambda\rbra{\bm{s},\bm{m}}}\ket{\alpha}_A = \frac{I+ \rbra{-1}^{m_t\oplus \lambda\rbra{\bm{s},\bm{m}}}P}{2}\ket{\alpha}_A = \frac{1+\rbra{-1}^{m_t\oplus \lambda\rbra{\bm{s},\bm{m}} \oplus \eta}}{2}\ket{\alpha}_A.
\end{equation*}
The only nonzero branch satisfies
\begin{equation*}
    m_t = \lambda\rbra{\bm{s},\bm{m}} \oplus \eta.
\end{equation*}
The planner emits a classical assignment instruction
\begin{equation*}
    \iDetMeas\rbra{m_t, \eta, \lambda\rbra{\bm{s},\bm{m}}},
\end{equation*}
which assigns $m_t = \lambda\rbra{\bm{s},\bm{m}} \oplus \eta$ from the currently available noise and measurement variables during sampling.
Neither a tableau update nor an update to the dense active-state vector is required.

\paragraph{Active-diagonal measurements.}
Suppose $d = 0$, $a = 0$ and $b \neq 0$.
Since $P$ is Hermitian, write
\begin{equation*}
    P\ket{\overline{x}}_A = \mu \overline{Z}^b_A\overline{Z}^c_D\ket{\overline{x}} = \rbra{-1}^{\eta}\overline{Z}^b_A\ket{\overline{x}}_A = \rbra{-1}^{b\cdot x \oplus \eta}\ket{\overline{x}}_A
\end{equation*}
with $\mu = \rbra{-1}^{\eta}$, $\eta \in \FF_2$.
Thus, the effective outcome $y_t$ retains the active-basis states $\ket{\overline{x}}_{A}$ for which
\begin{equation}\label{eq:active-diagonal-constraint}
    y_t = b\cdot x \oplus \eta.
\end{equation}
Choose an active pivot $h$, $1\leq h \leq k$, such that $P$ anticommutes with $\overline{X}_h$, equivalently $b_h = 1$.
Using the pivot conjugate $\overline{X}_{h}$ and $a,d=0$, the tableau update is
\begin{align*}
    \overline{Z}_j'
    &=
    \overline{Z}_j
    \overline{X}_{h}^{a_j} = \overline{Z}_j,
    &
    \overline{X}_j'
    &=
    \overline{X}_j
    \overline{X}_{h}^{b_j},
    &
    1\leq j \leq k,  j \neq h, \\
    \overline{Z}_{k+j}'
    &=
    \overline{Z}_{k+j}
    \overline{X}_{h}^{d_{j}} = \overline{Z}_{k+j},
    &
    \overline{X}_{k+j}'
    &=
    \overline{X}_{k+j}
    \overline{X}_{h}^{c_j},
    &
    1\leq j \leq n-k, \\
    \overline{Z}_h' &= \rbra{-1}^{\eta} P & \overline{X}_h' &= \overline{X}_h.
\end{align*}
These updates make every non-pivot tableau generator commute with $P$ while preserving the canonical stabilizer--destabilizer commutation relations.
Since $\rbra{-1}^{\eta}P = \overline{Z}^b\overline{Z}^c$ is in the stabilizer group generated by $\overline{Z}_1,\ldots \overline{Z}_n$,
the reference stabilizer state is unchanged:
\begin{equation*}
    \ket{\overline{0}}^{\text{new}} = \ket{\overline{0}}^{\text{old}}.
\end{equation*}
The induced basis change is
\begin{equation*}
    \ket{\overline{u}}^{\text{new}}_A = \prod_{j\neq h}
    \rbra*{\overline{X}_j\overline{X}_h^{b_j}}^{u_j} \overline{X}_h^{u_h}
    \ket{\overline{0}}^{\mathrm{old}} = \overline{X}_h^{\sum_{j\neq h}b_ju_j}\ket{\overline{u}}_{A}^{\text{old}}.
\end{equation*}
For $u \in \FF_2^k$, define $x\rbra{b,u} \in \FF_2^k$ by
\begin{equation*}
    {x_j\rbra{b,u}} = u_j \quad \rbra{j \neq h}, \qquad {x_h\rbra{b,u}} = u_h \oplus \bigoplus_{j \neq h} b_ju_j.
\end{equation*}
We have
\begin{equation*}
    \ket{\overline{u}}^{\text{new}}_A = \ket{\overline{x\rbra{b,u}}}_A^{\text{old}}.
\end{equation*}
Because $b_h = 1$, this map satisfies
\begin{equation*}
    b \cdot x\rbra{b,u}  = u_h.
\end{equation*}
Therefore, the branch condition \cref{eq:active-diagonal-constraint} becomes
\begin{equation*}
    u_h = y_t \oplus \eta,
\end{equation*}
which means the measurement projects the state into the $\rbra{-1}^{u_h}$ eigenspace of $\overline{Z}'_h$.
To store the post-measurement tableau in the $+1$ eigenspace of $\overline{Z}'_h$, the planner propagates the symbolic correction
\begin{equation*}
    \rbra*{\overline{X}_h'}^{u_h} = \rbra*{\overline{X}_h'}^{y_t \oplus \eta} = \rbra*{\overline{X}_h'}^{m_t\oplus \lambda\rbra{\bm{s},\bm{m}} \oplus \eta}
\end{equation*}
into the remaining planned operations.
The pivot pair $\rbra{\overline{Z}_h',\overline{X}_h'}$ is then moved to the dormant block, decreasing the active width by one.
The active generators with indices larger than $h$ are relabeled by shifting their indices down by one.

The planner emits
\begin{equation*}
    \iActiveDiagMeas\rbra*{m_t, b, h, \eta, \lambda\rbra*{\bm{s},\bm{m}}}.
\end{equation*}
For sampling, let $u(z,\delta)\in\FF_2^k$ denote the bit string obtained from $z\in\FF_2^{k-1}$ by inserting $\delta$ at position $h$:
\begin{equation*}
    u(z,\delta) = z_1\cdots z_{h-1}\,\delta\,z_h\cdots z_{k-1}.
\end{equation*}
The probability of recording $m_t \in \cbra{0,1}$, which implies $u_h = m_t \oplus \lambda\rbra{\bm{s},\bm{m}}\oplus \eta$, is
\begin{equation*}
    \Pr\rbra{m_t} = \sum_{z \in \FF_2^{k-1}} \abs*{\alpha_{x\rbra{b,u\rbra{z, m_t \oplus \lambda\rbra{\bm{s},\bm{m}}\oplus \eta}}}}^2.
\end{equation*}
After sampling $m_t$, the dense active-state vector is compacted as
\begin{equation*}
    \alpha_{z}' = \frac{\alpha_{x\rbra{b,u\rbra{z, m_t \oplus \lambda\rbra{\bm{s},\bm{m}}\oplus \eta}}}}{\sqrt{\Pr\rbra{m_t}}}, \qquad z \in \FF_2^{k-1}.
\end{equation*}

\paragraph{Active-nondiagonal measurements.}
Suppose $d = 0$ and $a \neq 0$.
By \cref{eq:active-logical-action}
\begin{equation*}
    P\ket{\overline{x}}_A = \mu \overline{X}^a\overline{Z}^b \overline{Z}^c \ket{\overline{x}}_A = \mu \rbra{-1}^{b\cdot x} \ket{\overline{x\oplus a}}_A.
\end{equation*}
Choose an active pivot $h$, $1\leq h\leq k$, such that $P$ anticommutes with $\overline{Z}_h$, equivalently $a_h = 1$.
Using the pivot conjugate $\overline{Z}_h$ and $d = 0$, the tableau update is
    \begin{align*}
        \overline{Z}_j'
        &=
        \overline{Z}_j
        \overline{Z}_{h}^{a_j},
        &
        \overline{X}_j'
        &=
        \overline{X}_j
        \overline{Z}_{h}^{b_j},
        &
        1\leq j \leq k,  j \neq h, \\
        \overline{Z}_{k+j}'
        &=
        \overline{Z}_{k+j}
        \overline{Z}_{h}^{d_{j}} = \overline{Z}_{k+j},
        &
        \overline{X}_{k+j}'
        &=
        \overline{X}_{k+j}
        \overline{Z}_{h}^{c_j},
        &
        1\leq j \leq n-k, \\
        \overline{Z}_h' &= P & \overline{X}_h' &= \overline{Z}_h.
    \end{align*}
These updates make every non-pivot tableau generator commute with $P$ and preserve the canonical stabilizer--destabilizer commutation relations.
The new active basis is
    \begin{align*}
        \ket{\overline{0}}^{\text{new}} &= \frac{I+P}{\sqrt{2}} \ket{\overline{0}}^{\text{old}} = \frac{1}{\sqrt{2}}\rbra*{\ket{\overline{0}}_A^{\text{old}}+\mu\ket{\overline{a}}_A^{\text{old}}}, \\
        \ket{\overline{x}}^{\text{new}}_A &= \rbra*{\overline{X}_1\overline{Z}_h^{b_1}}^{x_1}\cdots \rbra*{\overline{X}_{h-1}\overline{Z}_h^{b_{h-1}}}^{x_{h-1}} \overline{Z}_h^{x_h} \rbra*{\overline{X}_{h+1}\overline{Z}_h^{b_{h+1}}}^{x_{h+1}} \cdots \rbra*{\overline{X}_k\overline{Z}_h^{b_k}}^{x_k}\ket{\overline{0}}^{\text{new}} \\
        &= \rbra*{\overline{Z}_h^{x_h+\sum_{1\leq j \leq k, j\neq h}b_jx_j}} \overline{X}_1^{x_1}\cdots\overline{X}_{h-1}^{x_{h-1}} \overline{X}_{h+1}^{x_{h+1}} \cdots \overline{X}_{k}^{x_k}\ket{\overline{0}}^{\text{new}} \\
        &= \rbra*{\overline{Z}_h^{x_h+\sum_{1\leq j \leq k, j\neq h}b_jx_j}} \frac{1}{\sqrt{2}}\rbra*{\ket{\overline{x\sbra{h \mapsto 0}}}_A^{\text{old}}+\mu\ket{\overline{x\sbra{h \mapsto 0}\oplus a}}_A^{\text{old}}} \\
        &= \frac{1}{\sqrt{2}}\rbra*{\ket{\overline{x\sbra{h \mapsto 0}}}_A^{\text{old}}+\mu \rbra{-1}^{x_h+\sum_{1\leq j \leq k, j\neq h}b_jx_j}\ket{\overline{x\sbra{h \mapsto 0}\oplus a}}_A^{\text{old}}},
    \end{align*}
For $z \in \FF_2^{k-1}$ and $\delta \in \FF_2$, define $u\rbra{z, \delta}$ by
\begin{equation*}
    u\rbra{z,\delta} = z_1\cdots z_{h-1}\delta z_{h}\cdots z_{k-1}.
\end{equation*}
We have
\begin{equation*}
    \ket{\overline{u\rbra{z,\delta}}}_A^{\text{new}} = \frac{1}{\sqrt{2}}\rbra*{\ket{\overline{u\rbra{z,0}}}_A^{\text{old}}+\mu \rbra{-1}^{b\cdot u\rbra{z,0} \oplus \delta}\ket{\overline{u\rbra{z,0}\oplus a}}_A^{\text{old}}}.
\end{equation*}
Indeed,
\begin{equation*}
    P \ket{\overline{u\rbra{z,\delta}}}_A^{\text{new}} = \rbra{-1}^{\delta} \ket{\overline{u\rbra{z,\delta}}}_A^{\text{new}},
\end{equation*}
so the new pivot coordinate $\delta$ is precisely the eigenspace label of the measured Pauli $P$.

The inverse relations are
\begin{align*}
    \ket{\overline{u\rbra{z,0}}}_A^{\text{old}} & = \frac{1}{\sqrt{2}}\rbra*{\ket{\overline{u\rbra{z,0}}}_A^{\text{new}} + \ket{\overline{u\rbra{z,1}}}_A^{\text{new}}}, \\
    \ket{\overline{u\rbra{z,0}\oplus a}}_A^{\text{old}} &= \frac{\mu^*\rbra{-1}^{b\cdot u\rbra{z,0}}}{\sqrt{2}}\rbra*{\ket{\overline{u\rbra{z,0}}}_A^{\text{new}} - \ket{\overline{u\rbra{z,1}}}_A^{\text{new}}}.
\end{align*}
Thus, for each pair $\cbra*{u\rbra{z,0}, u\rbra{z,0}\oplus a}$,
\begin{align*}
    & \Pi_{P, y_t} \rbra*{\alpha_{u\rbra{z,0}} \ket{\overline{u\rbra{z,0}}}_A^{\text{old}} + \alpha_{u\rbra{z,0}\oplus a} \ket{\overline{u\rbra{z,0}\oplus a}}_A^{\text{old}}} \\
    ={}& \ketbra{\overline{u\rbra{z,y_t}}}{\overline{u\rbra{z,y_t}}}_A^{\text{new}} \rbra*{\alpha_{u\rbra{z,0}} \ket{\overline{u\rbra{z,0}}}_A^{\text{old}} + \alpha_{u\rbra{z,0}\oplus a} \ket{\overline{u\rbra{z,0}\oplus a}}_A^{\text{old}}}\\
    ={}& \beta_{z}\rbra*{y_t}\ket{\overline{u\rbra{z, y_t}}}_A^{\text{new}},
\end{align*}
where
\begin{equation}
    \beta_{z}\rbra*{y_t} = \frac{\alpha_{u\rbra{z,0}}+\mu^*\rbra{-1}^{b\cdot u\rbra{z,0}\oplus y_t} \alpha_{u\rbra{z,0}\oplus a}}{\sqrt{2}}.
\end{equation}
Thus, the projector produces a state in the $\rbra{-1}^{y_t}$ eigenspace of $\overline{Z}'_h$.
To store the post-measurement tableau in the $+1$ eigenspace of $P = \overline{Z}'_h$, the planner propagates the symbolic correction
\begin{equation*}
    \rbra*{\overline{X}_h'}^{y_t} = \rbra*{\overline{X}_h'}^{m_t\oplus \lambda\rbra{\bm{s},\bm{m}}}
\end{equation*}
into the remaining planned operations.
The pivot pair $\rbra{\overline{Z}_h',\overline{X}_h'}$ is then moved to the dormant block, decreasing the active width by one.

The planner emits
\begin{equation*}
    \iActivePairMeas\rbra*{m_t, a, b, h, \mu, \lambda\rbra*{\bm{s},\bm{m}}}.
\end{equation*}
During sampling, the probability of obtaining $m_t \in \cbra{0,1}$, which implies $y_t = m_t\oplus \lambda\rbra{\bm{s},\bm{m}}$, is
\begin{equation*}
    \Pr\rbra{m_t} = \sum_{z \in \FF_2^{k-1}} \abs*{\beta_z\rbra{m_t\oplus \lambda\rbra{\bm{s},\bm{m}}}}^2.
\end{equation*}
After sampling $m_t$, the dense active-state vector is compacted as
\begin{equation*}
    \alpha_{z}' \gets \frac{\beta_z\rbra{m_t\oplus \lambda\rbra{\bm{s},\bm{m}}}}{\sqrt{\Pr\rbra{m_t}}}, \qquad z \in \FF_2^{k-1}.
\end{equation*}

In both active-measurement cases, the tableau is updated only during planning.
During sampling, the emitted instruction samples or determines $m_t$ and then projects, normalizes, and compacts the dense active-state vector from dimension $2^k$ to $2^{k-1}$.

\section{Sampling Symbolic Signs and Measurement Outcomes}
\label{sec:sampling}

Factorization and planning produce an ordered sequence of sampling instructions.
Each instruction acts on the dense active-state vector, the symbolic bits, the measurement record, or some combination of them.
The stabilizer--destabilizer tableau used during planning is not needed during sampling.

The input to the sampler consists of:
\begin{itemize}
    \item the probability distributions of the independent stochastic Pauli noise variables $\bm{s}$;
    \item an ordered instruction stream
    \begin{equation*}
        \mathcal{I}
        =
        \rbra{\mathcal{I}_1,\mathcal{I}_2,\ldots,\mathcal{I}_L};
    \end{equation*}
    \item the affine Boolean expressions appearing in the instruction arguments;
    %\item detector and observable annotations, expressed as parities of recorded measurement outcomes.
\end{itemize}
The instruction stream contains the instructions
\begin{gather*}
    \iActiveDiagRot, \quad
    \iActivePairRot, \quad
    \iPromoteRot, \\
    \iRandomMeas, \quad
    \iDetMeas, \quad
    \iActiveDiagMeas, \quad
    \iActivePairMeas
\end{gather*}
derived in \cref{sec:planning}.
Their individual semantics were given there; we now explain how they are combined in a single sampling pass.

\paragraph{Symbol variable assignments.}
During sampling, the Boolean symbols fall into three classes.
\begin{itemize}
    \item Independent symbols include the noise variables $\bm{s}$ and the measurement symbols $\bm{m}_1$ produced by $\iRandomMeas$; they are sampled independently of the dense active-state vector.
    \item Deterministic measurement symbols $\bm{m}_2$ are assigned by $\iDetMeas$ from previously assigned symbols.
    \item Active-measurement symbols $\bm{m}_3$ are sampled by $\iActiveDiagMeas$ or $\iActivePairMeas$ from the current dense active-state vector.
\end{itemize}

A shot maintains a partial assignment
\begin{equation*}
    \nu:
    \bm{s}\cup\bm{m}_1\cup\bm{m}_2\cup\bm{m}_3
    \longrightarrow
    \FF_2
\end{equation*}
for these symbols.
For an affine Boolean expression $\lambda\rbra{\bm{s},\bm{m}}$, we write $\lambda[\nu]\in\FF_2$ for the value obtained by substituting the variables assigned so far.
The corresponding symbolic sign is then $\rbra{-1}^{\lambda[\nu]}$.

\paragraph{Single-shot sampling algorithm.}
At the beginning of each shot, the sampler initializes
\begin{equation*}
    k=0,
    \qquad
    \ket{\alpha}_A = 1,
    \qquad
    \bm{m}=\varnothing.
\end{equation*}
A shot then proceeds as follows:
\begin{enumerate}
    \item Sample the noise assignment $\bm{s}$ from the specified Pauli noise distributions and sample the symbols $\bm{m}_1$ independently and uniformly; store these values in $\nu$.

    \item Traverse the instruction stream
    $\mathcal{I}_1,\ldots,\mathcal{I}_L$ in order.

    \item For the current instruction $\mathcal{I}_\ell$, evaluate every affine Boolean expression in its arguments under the current assignment $\nu$.

    \item Execute the concrete instruction obtained after substitution. The instruction may update the dense active-state vector, change the active width $k$, and, for a measurement instruction, assign a new recorded bit $m_t$. Deterministic measurement instructions evaluate their emitted record expression. Active-measurement instructions compute the Born probabilities from the current vector, sample a branch, project and normalize the vector, and decrease the active width by one.

    \item Whenever a measurement bit $m_t$ is assigned, append it to the measurement record $\bm{m}$ and update the assignment $\nu$.
\end{enumerate}
All measurement-dependent affine Boolean expressions are causal: every measurement variable used by an instruction has already been assigned by an earlier instruction.
This property follows from the planning pass, which propagates symbolic corrections only into the remaining suffix of the instruction stream.
Consequently, sampling handles symbolic signs only by substitution:
\begin{equation*}
    \rbra{-1}^{\lambda\rbra{\bm{s},\bm{m}}}
    \quad\longmapsto\quad
    \rbra{-1}^{\lambda[\nu]}.
\end{equation*}
No Clifford frame is applied, no Pauli frame is updated, and no Pauli string is conjugated during sampling.

\paragraph{Sampling complexity.}
The same instruction stream is reused for all shots.
For each shot, the sampler draws the noise variables, evaluates affine Boolean expressions, records measurement outcomes, and executes the instructions that act on the dense active-state vector.

Let $n_t$ be the number of non-Clifford Pauli rotations; each emits at most one active or promoted rotation instruction.
Let $n_m$ be the number of Pauli measurements, and let $n_{m,\mathrm{active}}^{\mathrm S}$ be the number that emit an $\iActiveDiagMeas$ or $\iActivePairMeas$ instruction.
Let $n_e$ be the number of stochastic Pauli noise locations, and let $k_{\max}^{\mathrm S}$ be the peak active width, namely the maximum number of active stabilizer coordinates reached during a shot.

The state-vector part of sampling costs\footnote{In active rotations, the values of $b\cdot x$ for $x\in\FF_2^k$ can be incrementally computed with $O\rbra{2^k}$ cost.}
\begin{equation*}
    O\rbra*{\rbra*{n_t+n_{m,\mathrm{active}}^{\mathrm S}}
    2^{k_{\max}^{\mathrm S}}}.
\end{equation*}
This term includes every instruction that updates the dense active-state vector.
Dormant-nondiagonal random and dormant-diagonal deterministic measurements do not contribute to this state-vector cost.

It remains to account for the cost of evaluating symbolic signs.
The symbolic variables consist of measurement-record variables and variables associated with stochastic Pauli noise locations.
Assuming each noise location has a constant number of Pauli branches, the total number of symbolic variables is $V = O\rbra{n_m+n_e}$.
Let $\Lambda_{\mathrm{samp}}$ be the multiset of affine Boolean expressions evaluated during one shot.
These expressions define the symbolic signs attached to Pauli rotations, deterministic measurements, and active measurements.
Dormant-nondiagonal random measurements do not need to evaluate their symbolic signs in order to sample their recorded bit, since that bit is uniformly random.
For simplicity, we use the upper bound $\abs{\Lambda_{\mathrm{samp}}} = O\rbra{n_t+n_m}$.
For an affine Boolean expression $\lambda$, let $\wt\rbra{\lambda}$ denote the number of symbolic variables appearing in $\lambda$ after cancellation over $\FF_2$.
Define $d = 1 + \max_{\lambda\in\Lambda_{\mathrm{samp}}} \wt\rbra{\lambda}$.
The additive $1$ accounts for the constant term of an affine Boolean expression.
The cost of evaluating all symbolic signs in one shot is therefore
\begin{equation*}
    O\rbra*{\rbra*{n_t+n_m}d}.
\end{equation*}
In the worst case, an affine Boolean expression may contain all symbolic variables, so $d = O\rbra{n_m+n_e}$, and the symbolic-evaluation cost becomes
$O\rbra*{\rbra*{n_t+n_m}\rbra*{n_m+n_e}}$.
This worst-case bound is pessimistic for the fault-tolerant circuits considered by \symft{}.
\emph{In local fault-tolerant circuits, symbolic signs are typically sparse: a noise event or measurement branch often affects only a small spacetime neighborhood of later signs.
In this regime, $d$ is much smaller than $n_m+n_e$ and is often effectively constant.}

Combining the state-vector and symbolic-evaluation costs, the per-shot sampling complexity is
\begin{equation*}
    O\rbra*{\rbra*{n_t+n_{m,\mathrm{active}}^{\mathrm S}}
    2^{k_{\max}^{\mathrm S}} + \rbra*{n_t+n_m}d+n_e}.
\end{equation*}
The term $n_e$ accounts for sampling the stochastic Pauli noise choices.

\section{Implementation Design}
\label{sec:implementation}

We now describe how the preceding construction is implemented in \symft{}.
A central design choice is to move computations shared by all shots into the one-time planning pass.
Each shot then executes a compact sequence of instructions with little additional classical bookkeeping.

\paragraph{Commutation-aware simplification before planning.}
Before stabilizer-coordinate planning, \symft{} simplifies the factorized operator sequence using conservative local rewrites based on the commutation-aware rotation rules like \clifft{}~\cite{CL26}.
Between successive detector positions, rotations with the same Pauli body and compatible symbolic signs are fused across intervening commuting operations; the fused rotation is cancelled when the signed angles sum to zero.
The same pass may move a measurement earlier across commuting rotations, thereby shortening the lifetimes of active coordinates.
These rewrites preserve detector boundaries, measurement-record order, and classical assignment-before-use dependencies.
Moving measurements without removing any operation may reduce sampling throughput.
Unrestricted movement is therefore allowed only within a segment in which rotation fusion has already removed an operation.
In all other cases, a measurement moves only toward a rotation with the same Pauli body, which the planner can reduce to a branch phase.

\paragraph{Precompiled sampling instructions.}
During planning, \symft{} performs one-time computations that would otherwise be repeated for every shot.
These computations include conjugating Pauli operators through the Clifford and symbolic Pauli frames, decomposing them into active and dormant stabilizer coordinates, simplifying affine Boolean expressions over $\FF_2$, translating rotations and measurements into specialized sampling instructions, choosing pivots for tableau updates, and preparing detector and observable parity checks.
The result is an ordered sequence of precompiled sampling instructions together with a plan for evaluating their symbolic expressions.
Each dense-state instruction is represented by a compact descriptor containing Pauli masks, phase and angle data, and, when needed, a pivot.
The sampling kernel derives basis-dependent source indices and coefficients from this descriptor instead of storing them in tables.\footnote{Precomputing a full coefficient table for every active instruction can slightly improve sampling throughput by avoiding this arithmetic, but requires $O(2^k)$ additional memory per instruction at active width $k$.  The accumulated tables can therefore cause prohibitively large memory consumption during planning, so \symft{} instead retains only constant-size kernel descriptors.}
Consequently, the sampling loop neither interprets the original circuit, updates a stabilizer--destabilizer tableau, nor conjugates Pauli operators through Clifford gates.
It only executes the prepared instruction sequence.

\paragraph{Adaptive product-component representation.}
Once the initial instruction plan has been formed, a deterministic lowering pass identifies exact product structure in the active coefficient tensor.
The CPU samplers can then represent the state by one small dense vector for each independent component, without materializing the full tensor product.
Promoting a dormant rotation introduces a new one-coordinate component. An operation supported within a single component applies the usual dense update to that component alone.
If an operation spans several components, the sampler first merges only the affected components by an exact Kronecker product; the same dense update is then applied to the merged component.
The associated local masks, pivots, and merge schedules are computed once and shared by all shots.
This representation is selected only when a conservative cost model predicts a clear benefit. The model accounts for local vector work, merge traffic, component dispatch, and allocated component capacity.

\paragraph{Reusing the prepared state.}
Compilation produces a reusable sampler that contains the instruction and expression plans together with execution buffers sized according to the planned active-state capacities and symbol counts.
Retaining the active-state, scratch, symbol, and record buffers across chunks and sampling calls avoids repeated setup and allocation in steady-state sampling.
When several CPU threads are used, each independently processed chunk has a separate worker buffer.

\paragraph{Sampling independent symbols.}
Random symbols encode the stochastic choices made within a shot.
Some of them are independent of the current active state, including Pauli noise choices, readout flips when present, and the uniformly random bits produced by dormant-nondiagonal measurements.
Two optimizations reduce the cost of sampling these exogenous symbols.
First, following the low-entropy noise-sampling method of \stim{}~\cite{Gidney21}, the sampler uses a compact procedure when one is available instead of drawing a separate Bernoulli variable at every rare-event location.
For example, in a sparse-noise regime, the distances between rare events can be sampled from a geometric distribution without testing every no-error location.
Second, as in the batch symbolic samplers of
\symphase{}~\cite{FY24b} and \tsim{}~\cite{HLZ26}, the sampler presamples the independent symbols needed by a chunk of shots before executing that chunk.
Their values are stored as packed bit columns across shots.
For an affine Boolean expression $\lambda = \lambda_0 \oplus v_1 \oplus \cdots \oplus v_q$, its values over the chunk are obtained by XORing the packed columns for $v_1,\ldots,v_q$ and applying the constant bit $\lambda_0$.
During expression preparation, the presampled exogenous contribution is separated from the causal residual.
Repeated or closely related partial expressions are also shared rather than evaluated independently for every instruction.
As a result, machine-word operations evaluate much of the symbolic layer for many shots simultaneously.
Updating the active-state representation remains the only quantum-state operation performed during sampling.

\paragraph{Single-shot and batch sampling.}
\symft{} provides both single-shot and batch samplers.
Unless otherwise stated, the comparisons in this work are single-threaded.
Batching here means processing several shots together within one thread; it is not a form of multithreading.

The single-shot sampler processes one shot at a time: it reads the sampled symbols required by the shot, evaluates the remaining causal symbolic signs, records measurement outcomes, and updates a single dense active-state vector.

The batch sampler processes $B$ shots using the same instruction sequence.
Batching primarily benefits the classical symbolic layer. Assignments to independent symbols, measurement records, branch bits, and detector bits are represented by packed columns over the batch, allowing many affine Boolean expressions to be evaluated or reused with only a few machine-word operations.
Its role in updating the dense active-state vectors is more limited.
Each live shot retains its own vector because symbolic signs and measurement branches may differ across shots.
Accordingly, the batch sampler applies the same contiguous dense kernels to each shot while amortizing symbolic evaluation over the batch.

\paragraph{Dense active-state vector layout and SIMD.}
For each shot, the dense active-state vector is stored in separate real and imaginary arrays indexed by the active-basis state.
The batch sampler arranges these vectors in shot-major order.
Writing $S = 2^{k_{\max}^{\mathrm S}}$, the amplitude of shot $r$ at active-basis index $x$ is stored as
\begin{equation*}
    \alpha_{r,x}
    =
    \texttt{active\_re}\sbra{r\cdot S+x}
    +
    i\,\texttt{active\_im}\sbra{r\cdot S+x}.
\end{equation*}
For an instruction at active width $k$, only the first $2^k$ entries of each shot's stride are live.

This layout preserves the memory order used by the single-shot sampler for each dense active-state vector.
A standalone dense operation uses an outer loop over shots and an inner loop over contiguous active-basis amplitudes.
%Consecutive active rotations are grouped into short runs. Their packed signs are evaluated first, and the resulting loop order is shot, rotation instruction, and active amplitude.
%\color{blue}The vector for each shot therefore remains contiguous and can stay in the cache across neighboring rotations, instead of being revisited by a full-batch sweep for every instruction.

For nondiagonal active operations, efficient SIMD execution depends primarily on the choice of pivot.
Pair updates have the form $\rbra{x,x\oplus a}$.
Whenever possible, the planner selects the highest nonzero bit of $a$ as this pivot.
The pair structure within each pivot block is then regular: one side follows contiguous basis order, while the other lies in the opposite pivot half with only its lower bits permuted.
The kernel can consequently load contiguous groups from both sides, apply a fixed lane permutation when needed, perform the pair update, and store the results.
In the common high-pivot case, this procedure preserves the exact pair semantics without requiring a general gather--scatter operation.
The CPU implementation chooses the best available vectorized kernel while retaining a scalar fallback.

\paragraph{Batch size and cache behavior.}
The batch size is chosen adaptively from the peak active width determined during planning.
The dominant state-vector memory footprint is proportional to
$B\,2^{k_{\max}^{\mathrm S}}$, up to the factor for real and imaginary
arrays.
Increasing the batch size benefits the packed symbolic layer because more shots share expression evaluation and instruction traversal, but it also increases cache pressure on the dense active-state vectors.
The automatic choice therefore limits the state-vector footprint instead of using a fixed number of shots for every circuit.
The presampling chunk size is then chosen so that each chunk contains at least one complete batch.

\paragraph{Detector postselection.}
Many error-detection and fault-tolerant circuits reject a shot when a specified detector event occurs.
The sampler therefore evaluates each postselected detector as soon as the required measurement bits become available.

In the single-shot sampler, a rejecting detector terminates the shot immediately.
No subsequent symbolic or state-vector work is then required for that shot, so early rejection can directly reduce the sampling cost.

In the batch sampler, a live-shot mask marks rejected shots as inactive.
Dense instructions can skip these shots because every remaining live shot has a contiguous dense active-state vector.
Different shots may nevertheless be rejected at different points, while later instructions can still change the active width or produce measurement records for the survivors.
Rejected shots may initially remain masked.
When compaction becomes necessary or beneficial, the sampler packs the surviving active states together with only the measurement and symbolic columns still required.
This exact, adaptive compaction balances the cost of moving live data against the cost of carrying rejected shots through subsequent instructions.
Detector postselection therefore tends to provide a smaller speedup in the batch sampler than in the single-shot sampler, particularly when rejection times vary across the batch.

\paragraph{GPU acceleration.}
The CUDA backend retains the compiled instruction and expression
programs on the device and processes independent shots in launch-sized chunks. On the general active-state path, one CUDA thread block owns one shot: its threads cooperatively execute the planned dense-state updates, while the shot's amplitudes, scratch space, symbolic conditions, and measurement record remain in shared memory throughout the instruction stream. When the device-side expression-prepass mode is selected and both the peak active width and the symbolic state are sufficiently small, a separate scalar kernel instead assigns one shot to each CUDA thread, avoiding block-wide synchronization.

The backend can either sample exogenous variables and evaluate their affine expressions within each shot kernel, or presample packed expression values in a separate device-side pass; the faster choice depends on the circuit's expression structure and active width. Detector postselection terminates a shot as soon as a rejecting detector is recorded.
For the aggregate-count interface, the device returns only one discarded flag and one logical-error flag per shot, which are accumulated into counters on the host instead of materializing full detector records.
The active-state arithmetic type is selected at build time; we provide FP32 and FP64 builds and use FP64 for the precision-matched comparisons with \soft{}.

\section{Benchmark}
\label{sec:benchmark}

This section evaluates the sampling throughput of \symft{} in two regimes. \footnote{All test circuits and experimental configurations used in this section are available at: \url{https://github.com/haoliri0/SymFT_Test}.}
We first consider near-Clifford circuits, including the injection and cultivation stages of magic-state cultivation (MSC), coherent-noise surface-code circuits, and magic-state distillation.
We then consider pure-Clifford surface-code circuits.
The comparisons include \clifft{}~{\cite{CL26}} and \tsim{}~\cite{HLZ26}, the previous \soft{} implementation~\cite{LZZ+25}, and, for pure-Clifford circuits, \stim{}~\cite{Gidney21}.

As discussed in Section~\ref{sec:implementation}, \symft{} provides both single-shot and batch samplers.
The single-shot sampler evolves one shot at a time and can immediately terminate a rejected shot.
The batch sampler retains a separate active-state vector for each shot, but shares the planned instruction stream and evaluates independent symbols and affine signs as packed columns across shots.
Thus, batching amortizes the symbolic and dispatch overhead even on a single CPU thread; on the GPU, many shots are executed concurrently.
\emph{All single-core \symft{} results on CPU reported below use the batch sampler.}

\paragraph{Experimental setup.}
CPU measurements use one logical processor of a dual-socket Intel Xeon Gold 5218R machine; each socket has 20 physical cores, and the measured process is pinned to logical processor~0 with its SMT sibling idle.
The processor supports AVX-512 and has a 2.10\,GHz base clock.
Two 512-bit FMA pipelines give nominal per-core issue ceilings of 64 FP32 or 32 FP64 operations per cycle, corresponding at the base clock to 134.4\,GFLOP/s FP32 and 67.2\,GFLOP/s FP64.
These are instruction-level ceilings rather than measured sustained rates.
GPU measurements use a complete NVIDIA GeForce RTX~4090 with 24\,GB of device memory.
Its nominal stock-clock peaks are approximately 82.6\,TFLOP/s in FP32 and 1.29\,TFLOP/s in FP64, reflecting the device's $1{:}64$ FP64-to-FP32 throughput ratio~\cite{NVIDIA4090,NVIDIACUDABPG}.

\symft{} and \clifft{} use complex FP64 on the CPU. \tsim{} 0.1.5 was run with JAX x64 enabled. The \symft{} and \soft{} GPU results use FP64.

Each successful entry is the arithmetic mean of two sampling-only measurements of approximately 60\,s.
Compilation, planning, and JIT warmup are excluded; compilation exceeding 300\,s is reported as DNC.
The numerator is the number of attempted shots, before detector rejection.
All detectors are postselected in the near-Clifford circuits except MSC $d=7$, for which postselection is disabled; postselection is also disabled for the pure-Clifford circuits.
%The sampler invocation includes noise generation, quantum-state sampling, and the tool's normal result production, while separate Python postprocessing is excluded. Because the public APIs return different output forms---aggregate counters for \symft{} and full detector samples for some baselines---the results compare the tested sampling paths rather than isolated kernels with identical output widths.

\subsection{Near-Clifford results}
\zlabel{sec:benchmark-near-clifford}

\paragraph{Magic-state cultivation.}
The MSC workloads contain the injection and cultivation stages, but not the subsequent Clifford-only escape stage, and are adapted from the protocol in~\cite{GSJ24}.
The $d=3$ and $d=5$ circuits contain 15 and 42 qubits and have peak active widths $k_{\max}^{\mathrm S}=4$ and $10$, respectively.
\cref{tab:benchmark-msc-cpu,tab:benchmark-msc-gpu} report the circuit metadata together with the exact CPU and GPU rates.

\begin{table}[H]
    \centering
    \setlength{\tabcolsep}{3.5pt}
    \caption{Single-core attempted-shot throughput (shots/s) for MSC injection and cultivation.
    M and k denote $10^6$ and $10^3$.
    Ops is the number of operations after unrolling and excluding \texttt{TICK}, \texttt{DETECTOR}, and \texttt{OBSERVABLE\_INCLUDE}; Non-Cli counts non-Clifford rotations.
    The $d=3$ and $d=5$ rows postselect all detectors.
    The exploratory $d=7$ row is CPU-only and has postselection disabled.
    DNC means that compilation did not complete within 300\,s.}
    \label{tab:benchmark-msc-cpu}
    %\begin{adjustbox}{max width=\textwidth}
    \begin{tabular}{lrrrr|rrr}
        \toprule
        Circuit & Qubits & Ops & Non-Cli & $k_{\max}^{\mathrm S}$ & \tsim{} & \clifft{} & \symft{}\\
        \midrule
        MSC $d=3$ & 15 & 676 & 29 & 4
            & 39.92 & 502.3\,k & \textbf{1.762\,M}\\
        MSC $d=5$ & 42 & 4{,}379 & 91 & 10
            & DNC & 42.67\,k & \textbf{107.35\,k}\\
        MSC $d=7$ & 80 & 15{,}161 & 127 & 19
            & DNC & 25.23 & \textbf{47.09} \\
        \bottomrule
    \end{tabular}
    %\end{adjustbox}
    
\end{table}

\begin{table}[H]
    \centering
    \setlength{\tabcolsep}{3pt}
    \caption{GPU attempted-shot throughput (shots/s) for MSC injection and cultivation.
    \symft{} and \soft{} use FP64; the effective \tsim{} CUDA path retains FP32 and complex64 intermediates despite JAX x64 mode.
    M and k denote $10^6$ and $10^3$.
    Ops and Non-Cli use the definitions in \cref{tab:benchmark-msc-cpu}.
    Both rows postselect all detectors, and DNC denotes a 300\,s compilation timeout.}
    \label{tab:benchmark-msc-gpu}
    %\begin{adjustbox}{max width=\textwidth}
    \begin{tabular}{lrrrr|rrr}
        \toprule
        Circuit & Qubits & Ops & Non-Cli & $k_{\max}^{\mathrm S}$ & \soft{} FP64 & \tsim{} CUDA & \symft{} CUDA \\
        \midrule
        MSC $d=3$ & 15 & 676 & 29 & 4
            & 331.50\,k & 26.93\,k & \textbf{68.04\,M} \\
        MSC $d=5$ & 42 & 4{,}379 & 91 & 10
            & 5.03\,k & DNC & \textbf{2.61\,M} \\
        \bottomrule
    \end{tabular}
    %\end{adjustbox}
\end{table}

The exploratory $d=7$ circuit has 80 qubits and $k_{\max}^{\mathrm S}=19$.
On one CPU core, \symft{} samples it at 47.09 shots/s, $1.86\times$ the 25.23 shots/s of \clifft{}.
We have not generated enough shots to validate the full $d=7$ output distribution or its rare logical-error behavior, and therefore make no correctness claim for this workload.
Its circuit structure was constructed following the MSC architecture in~\cite{GSJ24}; we include it only as a performance stress test for a larger active subspace.
We disable detector postselection so that the reported rates are not dominated by early rejection.

\paragraph{Coherent noise and distillation.}
The coherent-noise circuits apply a small $R_Z(0.02)$ over-rotation together with circuit-level noise to distance-$d$ surface-code circuits for $r$ syndrome rounds.
Their peak active widths increase from $4$ for $(d,r)=(3,1)$ to $7$, $12$, and $22$ for $(3,3)$, $(5,1)$, and $(5,5)$, respectively.
\cref{tab:benchmark-coherent-cpu,tab:benchmark-coherent-gpu} report the CPU and effective-GPU sampling rates, respectively.
On one CPU core, \symft{} reaches 1.91\,M, 422.15\,k, 30.29\,k, and 31.93 shots/s for these four circuits.
Relative to \clifft{}, the corresponding nominal ratios are $2.76\times$, $3.96\times$, $4.84\times$, and $107.50\times$.

\tsim{} is much faster on the one-round coherent-noise circuits:
276.77\,M versus 1.91\,M shots/s for $d=3$, and 93.09\,M versus 30.29\,k shots/s for $d=5$.
In both cases, ZX simplification eliminates all residual component tensors and \tsim{} selects its \texttt{numpy-direct} path. Sampling then reduces to a direct host-side map from noise bits to detector and observable bits.
\symft{}, by contrast, still executes its general FP64 active-state, measurement, and postselection pipeline. When a GPU is requested, the same two \tsim{} cases also remain host \texttt{numpy-direct} executions.

For the distillation workload, \symft{} reaches 1.12\,M shots/s on one CPU core, $16.84\times$ \clifft{} and $173.37\times$ \tsim{}.
On the GPU, it reaches 23.57\,M shots/s, $17.9\times$ the 1.31\,M shots/s of \tsim{}.
The approximately $92.0\%$ discard rates agree across the tools.
\soft{} cannot compile the coherent-noise and distillation input syntax because its converter does not support the required $R_Z$ and $R_X$ instructions.

The FP64 \symft{} CUDA backend successfully runs every reported near-Clifford workload except coherent $d=5,r=5$: at $k_{\max}^{\mathrm S}=22$, the per-shot shared-memory requirement exceeds the RTX~4090 per-block limit.

\begin{table}[H]
    \centering
    \setlength{\tabcolsep}{3.5pt}
    \caption{Single-core attempted-shot throughput (shots/s) for the coherent-noise and distillation workloads.
    M and k denote $10^6$ and $10^3$; Ops and Non-Cli use the definitions in \cref{tab:benchmark-msc-cpu}.
    All rows postselect all detectors.
    DNC denotes a 300\,s compilation timeout.
    The coherent $d=3,r=3$ rates are not a controlled speedup comparison because the \clifft{} and \symft{} discard rates differ.}
    \label{tab:benchmark-coherent-cpu}
    %\begin{adjustbox}{max width=\textwidth}
    \begin{tabular}{lrrrr|rrr}
        \toprule
        Circuit & Qubits & Ops & Non-Cli & $k_{\max}^{\mathrm S}$ & \tsim{} & \clifft{} & \symft{} \\
        \midrule
        Coherent $d=3,r=1$ & 26 & 173 & 65 & 4
            & \textbf{276.77\,M} & 690.21\,k & 1.91\,M \\
        Coherent $d=3,r=3$ & 26 & 415 & 195 & 7
            & DNC & 81.32\,k & \textbf{422.15\,k} \\
        Coherent $d=5,r=1$ & 64 & 533 & 209 & 12
            & \textbf{93.09\,M} & 6.25\,k & 30.29\,k \\
        Coherent $d=5,r=5$ & 64 & 2{,}073 & 1{,}045 & 22
            & DNC & 0.271 & \textbf{31.93} \\
        Distillation & 85 & 1{,}163 & 10 & 5
            & 6.46\,k & 66.49\,k & \textbf{1.12\,M} \\
        \bottomrule
    \end{tabular}
    %\end{adjustbox}
\end{table}

\begin{table}[H]
    \centering
    %\small
    \setlength{\tabcolsep}{3pt}
    \caption{Effective-GPU attempted-shot throughput (shots/s) for the coherent-noise and distillation workloads.
    \symft{} uses FP64, and M and k denote $10^6$ and $10^3$.
    N/S marks syntax unsupported by the \soft{} converter, DNC denotes a 300\,s compilation timeout, and SMEM denotes the RTX~4090 per-block shared-memory limit.}
    \label{tab:benchmark-coherent-gpu}
    %\begin{adjustbox}{max width=\textwidth}
    \begin{tabular}{lrrrr|rrr}
        \toprule
        Circuit & Qubits & Ops & Non-Cli & $k_{\max}^{\mathrm S}$ & \soft{} FP64 & \tsim{} CUDA & \symft{} CUDA \\
        \midrule
        Coherent $d=3,r=1$ & 26 & 173 & 65 & 4
            & N/S & \textbf{370.12\,M} & 145.16\,M \\
        Coherent $d=3,r=3$ & 26 & 415 & 195 & 7
            & N/S & DNC & \textbf{15.93\,M} \\
        Coherent $d=5,r=1$ & 64 & 533 & 209 & 12
            & N/S & \textbf{90.91\,M} & 480.30\,k \\
        Coherent $d=5,r=5$ & 64 & 2{,}073 & 1{,}045 & 22
            & N/S & DNC & SMEM \\
        Distillation & 85 & 1{,}163 & 10 & 5
            & N/S & 1.31\,M & \textbf{23.57\,M} \\
        \bottomrule
    \end{tabular}
    %\end{adjustbox}
\end{table}

\subsection{Pure-Clifford results}
\zlabel{sec:benchmark-pure-clifford}

The pure-Clifford workloads are rotated surface-code memory circuits at physical noise strength $p=10^{-3}$.
The $d=7,r=7$ circuit contains 97 qubits and 336 detectors, while the $d=9,r=9$ circuit contains 161 qubits and 720 detectors.
For these circuits $k_{\max}^{\mathrm S}=0$, so \symft{} performs only the symbolic and detector-processing portion of its sampling pipeline.
No detector is postselected.

\cref{tab:benchmark-pure-clifford-cpu} shows that \symft{} is the fastest measured single-core implementation on both circuits.
It reaches 2.06\,M shots/s at $d=7$, $2.52\times$ \stim{}, $22.05\times$ \clifft{}, and $1.68\times$ \tsim{}.
At $d=9$, it reaches 899.20\,k shots/s, $2.56\times$ \stim{}, $21.42\times$ \clifft{}, and $2.74\times$ \tsim{}.

\begin{table}[H]
    \centering
    \small
    \setlength{\tabcolsep}{6pt}
    \caption{Single-core attempted-shot throughput (shots/s) for pure-Clifford surface-code circuits.
    All measurements use one pinned CPU core.
    M and k denote $10^6$ and $10^3$.}
    \label{tab:benchmark-pure-clifford-cpu}
    %\begin{adjustbox}{max width=\textwidth}
    \begin{tabular}{lrrr|rrr}
        \toprule
        Circuit & Qubits & Ops & \stim{} & \clifft{} & \tsim{} & \symft{} \\
        \midrule
        Surface $d=7,r=7$ & 97 & 4{,}667
            & 816.93\,k & 93.41\,k & 1.22\,M & \textbf{2.06\,M} \\
        Surface $d=9,r=9$ & 161 & 9{,}997
            & 350.34\,k & 41.97\,k & 327.59\,k & \textbf{899.20\,k} \\
        \bottomrule
    \end{tabular}
    %\end{adjustbox}
\end{table}

\subsection{Differences between \symft{} and \soft{}}
\label{sec:benchmark-symft-soft}

The principal source of the sampling-throughput difference between \symft{} and its predecessor \soft{} is the treatment of shared Clifford evolution.
Whereas \soft{} evolves the full trajectory for every shot with a private generalized-stabilizer tableau, coefficient map, and noise history~\cite{LZZ+25}, \symft{} compiles the deterministic Clifford evolution once and shares its symbolic Clifford--Pauli frame and stabilizer-coordinate plan across shots.
Each shot then updates only active coefficients, symbolic signs, measurement records, and detector state.
The $205\times$ and $518\times$ FP64 GPU ratios for MSC $d=3$ and $d=5$ in \cref{tab:benchmark-msc-gpu} primarily reflect this removal of per-shot tableau evolution.

This optimization changes the capability tradeoff.
\soft{}'s retained trajectories make dynamic detector-error-model construction and online decoding natural.
\symft{} reduces repeated work by moving the trajectory offline, but supporting those features would require extending its compiled symbolic interface.

When \soft{} was developed, 16 NVIDIA H800 GPUs were sufficient for the target MSC simulation, so this separation was not pursued~\cite{LZZ+25}.
In retrospect, a global tableau trajectory plus per-shot fault- and branch-dependent phases would have sufficed.
\symft{} adopts this lesson by combining \clifft{}'s active-subspace view~\cite{CL26} with \symphase{}'s batched symbolic phases~\cite{FY24b}, stabilizer coordinates, and packed SIMD and GPU kernels.
This combination produces the throughput gains above while preserving the exact noisy, adaptive sampling semantics.

\section{Related Work}
\label{sec:related-work}

\paragraph{Stabilizer-circuit simulation.}
The Gottesman--Knill theorem permits efficient classical simulation of Clifford circuits with stabilizer preparations and Pauli measurements~\cite{Gottesman98}.
Aaronson and Gottesman introduced the stabilizer--destabilizer tableau and an efficient measurement-update algorithm~\cite{AG04},
while Anders and Briegel developed a graph-state representation with favorable performance for many stabilizer circuits~\cite{AB06}.
Other stabilizer-state representations include quadratic-form expansions~\cite{BH22}.
\stim{} combines an inverse stabilizer tableau, cache-efficient bit-packed data structures, SIMD kernels, and a compile-once sampling procedure based on a reference sample and batched Pauli frames~\cite{Gidney21}.
\symphase{} instead represents Pauli faults and random-measurement outcomes by affine Boolean expressions in the phase columns of a tableau, allowing many samples to be obtained by substitution after a single circuit traversal~\cite{FY24b}.
Adjoint-based code simulation removes per-shot fault propagation in Clifford circuits by precomputing the backward propagation of a small set of Pauli observables~\cite{DP23}.
For stabilizer circuits under Pauli noise, treating each measurement outcome as a detector makes \stim{}'s detector error model, adjoint back-propagation, and \symphase{}'s forward expressions equivalent encodings of the fault-to-measurement map.
Direct sampling then reduces to sparse binary matrix--vector multiplication, which can be faster than \stim{}'s default Pauli-frame propagation when this map is sufficiently sparse.
A complementary noisy-stabilizer framework moves shared circuit work into preprocessing by absorbing Clifford gates and compatible Pauli measurements into a reference stabilizer state. The remaining Pauli-diagonal noise channels can then be used for exact Pauli-expectation evaluation or lower-cost Pauli-frame sampling~\cite{AMM26}.
Recent GPU-oriented stabilizer simulators include STABSim~\cite{GLW+25} and QuaSARQ~\cite{OTL26}.

\paragraph{Closest universal QEC simulators.}
\clifft{}~\cite{CL26} and \tsim{}~\cite{HLZ26} are the closest independent general simulators to \symft{}.
Our previous simulator \soft{}~\cite{LZZ+25} is \symft{}'s direct predecessor.
These methods place the exponential cost in different objects.
The comparison below concerns per-shot sampling costs and omits one-time compilation costs.
To keep notation uniform, let $n$ be the number of physical qubits, $n_t$ the number of non-Clifford Pauli rotations ($T/T^\dagger$ gates in the Clifford+$T$ setting), $n_m$ the number of Pauli measurements, and $n_e$ the number of stochastic Pauli noise locations.
\begin{itemize}
    \item \clifft{} stores the non-stabilizer part in a dynamically sized dense active-state vector and uses Clifford localization to reduce active Pauli operations to one virtual qubit~\cite{CL26}.
    Let $n_{m,\mathrm{active}}^{\mathrm C}$ be the number of measurements that act on \clifft{}'s active subspace, and let $k_{\max}^{\mathrm C}$ be its peak active width, measured in active virtual qubits.
    Its worst-case per-shot cost is
    \begin{equation*}
        O\rbra*{(n_t+n_{m,\mathrm{active}}^{\mathrm C})
        2^{k_{\max}^{\mathrm C}} +(n_t+n_m+n_e)n}.
    \end{equation*}
    Its Clifford frame, localization choices, and active-set trajectory are shared across shots. However, each shot still updates an $n$-qubit Pauli frame, and localization that touches active virtual qubits also produces Clifford operations on the dense active-state vector.
    In \symft{}, let $n_{m,\mathrm{active}}^{\mathrm S}$ be the number of active measurements, $k_{\max}^{\mathrm S}$ be its peak active width in stabilizer coordinates, and $d$ be one plus the maximum number of variables in an affine Boolean expression evaluated during sampling.
    Its per-shot cost bound is
    \begin{equation*}
        O\rbra*{\rbra*{n_t+n_{m,\mathrm{active}}^{\mathrm S}} 2^{k_{\max}^{\mathrm S}} + \rbra*{n_t+n_m}d+n_e}.
    \end{equation*}
    Both methods therefore use the same type of representation-level parameter: peak active width.
    The two active widths will coincide at every step, i.e., $k_{\max}^{\mathrm S}=k_{\max}^{\mathrm C}$, whenever both planners process the same residual Pauli sequence and maintain maximum-rank dormant stabilizer subgroups after every prefix.
    This equality is conditional: reordering, fusion, or different coordinate-update choices can change the compiled trajectory, so we retain distinct superscripts.
    Thus, \symft{} does not necessarily have a smaller exponential parameter.
    Its direct updates in stabilizer coordinates avoid localization-induced dense Clifford sweeps, and its symbolic layer replaces \clifft{}'s online $n$-bit Pauli-frame work with sign evaluation costing $O(d)$.
    For local fault-tolerant circuits with sparse symbolic signs, \symft{} replaces the physical-qubit factor $n$ by the sparsity parameter $d$.
    \item \tsim{} maps Pauli noise to Boolean parameters in a ZX diagram, performs global ZX reduction, decomposes the residual non-Clifford diagram into $\chi$ stabilizer terms, and compiles the resulting tensors for vectorized CPU or GPU evaluation~\cite{HLZ26}.
    For Clifford+$T$ circuits, the partial Cat5 decomposition gives $\chi=O(2^{0.396n_T})$, where $n_T$ is the number of $T/T^\dagger$ gates~\cite{KWV22}.
    \tsim{} generalizes this decomposition to arbitrary rotation angles and reproduces the same asymptotic exponent~\cite{HLZ26}; global diagram simplification can make the post-reduction value of $\chi$ substantially smaller.
    Its per-shot cost bound is
    \begin{equation*}
        O\rbra*{n_e + \chi n_t^2}.
    \end{equation*}
    This bound assumes that ZX reduction separates detector components from observable components and that the number of observables is $O(1)$. It does not cover raw measurement sampling, which can require $n_m$ autoregressive marginals.
    The cost of \tsim{} is governed by the global post-reduction term count $\chi$, whereas that of \symft{} is governed by the local active width $k_{\max}^{\mathrm S}$, which can decrease after a measurement.
    Neither parameter dominates the other: ZX simplification can expose global cancellations that local planning misses, while a dense active-state representation can discard non-stabilizer degrees of freedom created by earlier non-Clifford gates.
    \item \soft{} assigns each shot its own $n$-qubit generalized-stabilizer tableau and sparse coefficient map~\cite{LZZ+25}.
    Let $r$ be the current number of nonzero coefficients.  
    A local Clifford gate costs $O(n)$, whereas a Pauli measurement or $T/T^\dagger$ gate costs $O(rn+n^2)$. A measurement can decrease $r$, and a $T$ gate can at most double it.
    Let $n_c$ count the local Clifford gates and let $r_{\max}$ be the peak sparse support. 
    Treating each stochastic Pauli noise location as a possible local Pauli update gives the worst-case quantum-state update bound
    \begin{equation*}
        O\rbra*{(n_c+n_e)n+(n_m+n_t)(r_{\max}n+n^2)}.
    \end{equation*}
    This expression excludes random-number generation, detector and postselection processing, and other classical overheads that are not specified separately.
    In the worst case, \(r_{\max}\leq 2^{k_{\max}^{\mathrm S}}\), and code constraints can make it much smaller.
    \symft{} shares the tableau trajectory across shots, removing \soft{}'s per-shot tableau updates.
    It does not uniformly dominate sparse storage: \soft{} can be preferable when \(r_{\max}\ll 2^{k_{\max}^{\mathrm S}}\), whereas \symft{} favors contiguous kernels on the dense active-state vector when the active subspace is compact.
\end{itemize}

\paragraph{Generalized-stabilizer representations.}
Stabilizer frames represent a state as a superposition of related stabilizer states with shared data~\cite{GMC13,GM15}.
The generalized-stabilizer formalism~\cite{Yoder12} makes this sharing explicit: a stabilizer--destabilizer tableau defines the orthonormal basis $\{\overline{X}_1^{\alpha_1}\cdots\overline{X}_n^{\alpha_n}\ket{\overline{0}}\}_{\alpha\in\FF_2^n}$, and one coefficient vector represents an arbitrary state.
\soft{} stores this vector sparsely for each shot~\cite{LZZ+25}, while the Pauli Frame Sparse Representation augments a similar sparse basis with Pauli histories that track relative phases~\cite{TA26}.
Stabilizer tensor networks use the same tableau-defined basis but encode its coefficient tensor as a tensor network~\cite{MG24,NHW25,CRG+26}.
Their cost is therefore governed by contraction or bond dimension rather than by the coefficient count alone.
Other works use the terms generalized stabilizer and extended stabilizer for different algebraic constructions, including generalized stabilizing sets and high-rank generators represented by Pauli sums~\cite{DD24,HDD+26}.
\symft{} instead moves the deterministic tableau trajectory and basis updates into its one-time planning pass; during sampling, each shot updates only the dense active-state vector.

\paragraph{Stabilizer rank and ZX-calculus.}
A separate approach decomposes a non-stabilizer state or operation into $\chi$ stabilizer terms.
Its exact or approximate cost is controlled by stabilizer rank, extent, Gauss-sum rank, or related magic monotones, and is typically exponential in a global amount of magic~\cite{BG16,BBC+19,MT24,HL19,QPG21,SRP+21,PRK22,Kocia22}.
Pauli-based computation provides an equivalent universal model based on adaptive Pauli measurements of magic-state registers~\cite{BSS16}.
Stabilizer decompositions are also a key ingredient of later ZX-based simulators. 
ZX rewriting first exposes diagram-dependent structure; sum-over-stabilizer, partial and graphical decompositions, and procedural cuts then exploit that structure~\cite{KW22,KWV22,SK24,SK25}.
In particular, parametric rewriting can reuse reductions across parameter values~\cite{SK25}.
This global strategy underlies \tsim{} and is complementary to the local stabilizer-coordinate planning used by \symft{}.

\paragraph{Symbolic methods.}
Symbolic Pauli gates were introduced to isolate subcircuits during Clifford optimization~\cite{BSH+21}, and Pauli-rotation representations similarly factor Clifford structure from non-Clifford phases~\cite{ZC19}.
\symphase{} specializes this idea to affine Boolean expressions in tableau phase columns, enabling many samples after a single stabilizer traversal~\cite{FY24b,FY24a}.
Parameterized ZX rewriting and \tsim{} share reduced diagrams and Clifford terms across Boolean noise assignments~\cite{SK25,HLZ26}.
\syqma{} instead introduces auxiliary qubits and a modified trace to obtain exact symbolic probabilities and logical error expressions with polynomial representation memory, but its evaluation time is exponential in the number of non-Clifford rotations and deterministic measurements~\cite{UA26}.
\symft{} places symbolic signs on the pulled-back rotations and projectors, so each shot evaluates affine Boolean expressions instead of updating Clifford and Pauli frames during sampling.
The symbolic signs of \symft{} could also be paired with a stabilizer-rank decomposition of the residual non-Clifford core, just as parameterized ZX methods combine symbolic parameters with sum-over-Clifford evaluation. We leave this hybrid approach to future work.

\paragraph{Pauli propagation.}
Pauli-propagation methods evolve Pauli expansions of observables or channels, usually in the Heisenberg picture, with cost controlled by the number of retained Pauli terms or propagation paths~\cite{RLC19}.
Noise-induced suppression of Pauli paths has led, under different assumptions, to polynomial-time approximation algorithms for random, variational, and more general noisy circuits~\cite{AGL+23,SWC+24,FRD+25,SYG+25,GCT25,AMR26}.
Rudolph et al.\ develop Pauli propagation from bit-level implementations and a tree-search formulation to circuit- and system-level applications, and provide PauliPropagation.jl as an end-to-end Julia implementation~\cite{RJT+26}.
At the level of reusable Pauli arithmetic, PauliEngine supports numerical and symbolic coefficients~\cite{MBK26}, while PauLIB emphasizes sorted arrays, a structure-of-arrays layout, and SIMD-parallel processing of Pauli strings and sums~\cite{K26}.
Although \symft{} and Pauli propagation both use Pauli algebra, they employ different representations: generalized-stabilizer simulation evolves Schr\"odinger-picture amplitudes in a tableau-defined basis, whereas Pauli propagation evolves operator or channel expansions in the Heisenberg picture and becomes approximate when those expansions are truncated.
Recent QEC work approximately simulates weak non-Pauli noise in Clifford circuits~\cite{MOH+25,HOR+26}, with related detector-model reductions building on the framework of~\cite{DTB+25}.

\paragraph{Rare-event sampling and protocol-specific simulation.}
Rare-event estimators reduce the number of shots needed to resolve small logical-failure probabilities through splitting~\cite{BV13,MGO25}, dynamical subset sampling~\cite{HWR24}, failure-spectrum and multi-seeded splitting methods~\cite{BCC+25}, and stratified fault injection with extrapolation~\cite{YP26}.
These methods are complementary to \symft{}: they reduce the number of required shots, whereas \symft{} reduces the cost of each exact shot.

Other methods exploit the structure of a particular fault-tolerant protocol, for example by reducing the relevant cultivation states to small Clifford sums~\cite{WZZ26}.
For circuit families in which every circuit-level Pauli fault propagates to a Clifford error at the end of the circuit, the sampling cost scales polynomially with both the number of qubits and the stabilizer or Pauli rank of the target logical state~\cite{SDK26}.
These results exploit the structure of particular protocols, whereas \symft{} addresses general high-throughput sampling of noisy, adaptive Clifford-dominated circuits.

\section{Conclusion and Outlook}
\label{sec:conclusion}

We introduced \symft{}, an exact simulator for noisy, adaptive Clifford-dominated circuits.
Its symbolic Clifford--Pauli frame factorization processes the Clifford evolution once and confines the effects of Pauli noise and feedback to affine signs on pulled-back rotations and projectors.
Its adaptive stabilizer-coordinate planner shares a tableau trajectory across shots and compiles direct sampling instructions for a dynamically sized active state.
Sampling therefore neither replays the original circuit nor reconstructs the tableau, and its dominant exponential cost is governed by the planned active structure rather than by the number of physical qubits.
Our C\texttt{++} CPU and CUDA GPU backends, provided through a Python package, outperform the compared baselines on the reported pure-Clifford QEC and magic-state-cultivation workloads.

Several directions could further improve the method.
First, the current implementation already applies a conservative commutation-aware optimization to the factorized operator sequence before planning, including compatible Pauli-rotation fusion and cancellation and guarded movement of commuting measurements, as described in \cref{sec:implementation}.
This realizes part of the sequence optimization used in \clifft{}~\cite{CL26}, but more comprehensive, cost-aware rewrites remain possible.
For example, Pauli-rotation $T$-count optimization~\cite{ZC19}, moving non-Clifford rotations later, and broader measurement scheduling could further reduce the number and lifetime of active operations.
Because a lower $T$ count does not by itself guarantee a lower peak active width, future optimizers should jointly consider $k_{\max}^{\mathrm S}$, symbolic-evaluation cost, and the total dense-vector work along the planned trajectory.
Second, the current product-component backend exploits exact separability but eagerly merges components when an operation couples them and does not attempt to split a merged component again.
One possible extension is to detect, at low cost, new exact factorizations created by later projections or cancellations.
More expressive structured representations could also retain partial rather than complete factorization.
For example, a tensor network over the same stabilizer coordinates could exploit a low but nonzero bond dimension in the coefficient tensor, connecting \symft{} with stabilizer tensor networks~\cite{MG24,NHW25,CRG+26}.
More generally, an adaptive sampler could choose among monolithic dense, exact product-component, sparse, and tensor-network representations according to their estimated costs.
Third, the symbolic sign layer could be combined with stabilizer-rank or ZX-based decompositions of the residual non-Clifford core, allowing one decomposition or reduced diagram to be reused across many noise and measurement assignments~\cite{BG16,KWV22,SK25,HLZ26}.
Retaining symbolic forms beyond per-shot substitution could also support parameter-dependent probabilities, logical-error expressions, and noise-parameter sweeps, in the spirit of \syqma{}~\cite{UA26} and analytical or compressed noisy-stabilizer simulation~\cite{AMM26}.
Finally, combining \symft{} with rare-event estimators could reduce both the cost of each shot and the number of shots needed to estimate very small logical-failure probabilities~\cite{BV13,HWR24,MGO25,BCC+25,YP26}.

\section*{Acknowledgement}

The authors used ChatGPT Pro (GPT-5.5/5.6) and OpenAI Codex for implementation, exploratory coding, preliminary literature searches, and language editing.
The authors verified all results, revised the manuscript as needed, and take full responsibility for its content.

\bibliographystyle{quantum}
\bibliography{main}

@article{Gidney21,
    doi = {10.22331/q-2021-07-06-497},
    url = {https://doi.org/10.22331/q-2021-07-06-497},
    title = {Stim: a fast stabilizer circuit simulator},
    author = {Gidney, Craig},
    journal = {{Quantum}},
    issn = {2521-327X},
    publisher = {{Verein zur F{\"{o}}rderung des Open Access Publizierens
                in den Quantenwissenschaften}},
    volume = 5,
    pages = 497,
    month = jul,
    year = 2021
}

@misc{HLZ26,
    title={Tsim: Fast Universal Simulator for Quantum Error Correction}, 
    author={Rafael Haenel and Xiuzhe Luo and Chen Zhao},
    year={2026},
    eprint={2604.01059},
    archivePrefix={arXiv},
    primaryClass={quant-ph},
    url={https://arxiv.org/abs/2604.01059}, 
}

@misc{CL26,
    title={Clifft: Fast Exact Simulation of Near-Clifford Quantum Circuits}, 
    author={Bradley A. Chase and Farrokh Labib},
    year={2026},
    eprint={2604.27058},
    archivePrefix={arXiv},
    primaryClass={quant-ph},
    url={https://arxiv.org/abs/2604.27058}, 
}

@misc{LZZ+25,
    title={{SOFT}: a high-performance simulator for universal fault-tolerant quantum circuits}, 
    author={Riling Li and Keli Zheng and Yiming Zhang and Huazhe Lou and Shenggang Ying and Ke Liu and Xiaoming Sun},
    year={2025},
    eprint={2512.23037},
    archivePrefix={arXiv},
    primaryClass={quant-ph},
    url={https://arxiv.org/abs/2512.23037}, 
}

@misc{NVIDIA4090,
    author={{NVIDIA Corporation}},
    title={{GeForce RTX 4090} specifications},
    year={2026},
    howpublished={\url{https://www.nvidia.com/en-us/geforce/graphics-cards/40-series/rtx-4090/}},
    note={Accessed 2026-07-19}
}

@manual{NVIDIACUDABPG,
    author={{NVIDIA Corporation}},
    title={{CUDA C++ Best Practices Guide}: instruction throughput},
    year={2026},
    url={https://docs.nvidia.com/cuda/cuda-c-best-practices-guide/},
    note={Accessed 2026-07-19}
}

@misc{TA26,
    title={Computing logical error thresholds with the Pauli Frame Sparse Representation},
    author={Thomas Tuloup and Thomas Ayral},
    year={2026},
    eprint={2603.14670},
    archivePrefix={arXiv},
    primaryClass={quant-ph},
    url={https://arxiv.org/abs/2603.14670},
}

@inproceedings{FY24b,
author = {Fang, Wang and Ying, Mingsheng},
title = {{SymPhase}: Phase Symbolization for Fast Simulation of Stabilizer Circuits},
year = {2024},
isbn = {9798400706011},
url = {https://doi.org/10.1145/3649329.3655902},
doi = {10.1145/3649329.3655902},
booktitle = {Proceedings of the 61st ACM/IEEE Design Automation Conference},
articleno = {32},
numpages = {6},
series = {DAC '24}
}

@article{FY24a,
author = {Fang, Wang and Ying, Mingsheng},
title = {Symbolic Execution for Quantum Error Correction Programs},
year = {2024},
issue_date = {June 2024},
volume = {8},
number = {PLDI},
url = {https://doi.org/10.1145/3656419},
doi = {10.1145/3656419},
journal = {Proc. ACM Program. Lang.},
month = jun,
articleno = {189},
numpages = {26}
}

@INPROCEEDINGS{Shor96,
  author={Shor, Peter W.},
  booktitle={Proceedings of 37th Conference on Foundations of Computer Science}, 
  title={Fault-tolerant quantum computation}, 
  year={1996},
  pages={56--65},
  doi={10.1109/SFCS.1996.548464}
}

@phdthesis{Gottesman97,
  author={Daniel Gottesman},
  title={Stabilizer Codes and Quantum Error Correction},
  school={California Institute of Technology},
  year={1997},
  eprint={quant-ph/9705052},
  archivePrefix={arXiv},
  primaryClass={quant-ph},
  url={https://arxiv.org/abs/quant-ph/9705052},
}

@misc{Gottesman98,
      title={The {H}eisenberg Representation of Quantum Computers}, 
      author={Daniel Gottesman},
      year={1998},
      eprint={quant-ph/9807006},
      archivePrefix={arXiv},
      primaryClass={quant-ph},
      url={https://arxiv.org/abs/quant-ph/9807006}, 
}

@article{BK05,
  title = {Universal quantum computation with ideal {C}lifford gates and noisy ancillas},
  author = {Bravyi, Sergey and Kitaev, Alexei},
  journal = {Phys. Rev. A},
  volume = {71},
  issue = {2},
  pages = {022316},
  numpages = {14},
  year = {2005},
  month = {Feb},
  publisher = {American Physical Society},
  doi = {10.1103/PhysRevA.71.022316},
  url = {https://link.aps.org/doi/10.1103/PhysRevA.71.022316}
}

@article{GF19,
  doi = {10.22331/q-2019-04-30-135},
  url = {https://doi.org/10.22331/q-2019-04-30-135},
  title = {Efficient magic state factories with a catalyzed {$|CCZ\rangle$} to {$2|T\rangle$} transformation},
  author = {Gidney, Craig and Fowler, Austin G.},
  journal = {{Quantum}},
  issn = {2521-327X},
  publisher = {{Verein zur F{\"{o}}rderung des Open Access Publizierens in den Quantenwissenschaften}},
  volume = {3},
  pages = {135},
  month = apr,
  year = {2019}
}

@article{Litinski19,
  doi = {10.22331/q-2019-12-02-205},
  url = {https://doi.org/10.22331/q-2019-12-02-205},
  title = {Magic {S}tate {D}istillation: {N}ot as {C}ostly as {Y}ou {T}hink},
  author = {Litinski, Daniel},
  journal = {{Quantum}},
  issn = {2521-327X},
  publisher = {{Verein zur F{\"{o}}rderung des Open Access Publizierens in den Quantenwissenschaften}},
  volume = {3},
  pages = {205},
  month = dec,
  year = {2019}
}

@INPROCEEDINGS{HIF24,
  author={Hirano, Yutaka and Itogawa, Tomohiro and Fujii, Keisuke},
  booktitle={2024 IEEE International Conference on Quantum Computing and Engineering (QCE)}, 
  title={Leveraging Zero-Level Distillation to Generate High-Fidelity Magic States}, 
  year={2024},
  volume={01},
  number={},
  pages={843--853},
  doi={10.1109/QCE60285.2024.00104}
}

@article{WHY24,
  title={Constant-overhead magic state distillation},
  author={Adam Wills and Min-Hsiu Hsieh and Hayata Yamasaki},
  journal={Nature Physics},
  year={2024},
  volume={21},
  pages={1842--1846},
  doi={10.1038/s41567-025-03026-0}
}

@article{LTF+25,
  title = {Low-Overhead Magic State Distillation with Color Codes},
  author = {Lee, Seok-Hyung and Thomsen, Felix and Fazio, Nicholas and Brown, Benjamin J. and Bartlett, Stephen D.},
  journal = {PRX Quantum},
  volume = {6},
  issue = {3},
  pages = {030317},
  numpages = {50},
  year = {2025},
  month = {Jul},
  publisher = {American Physical Society},
  doi = {10.1103/ch5r-cnfq},
  url = {https://link.aps.org/doi/10.1103/ch5r-cnfq}
}

@article{ITH+25,
  title = {Efficient Magic State Distillation by Zero-Level Distillation},
  author = {Itogawa, Tomohiro and Takada, Yugo and Hirano, Yutaka and Fujii, Keisuke},
  journal = {PRX Quantum},
  volume = {6},
  issue = {2},
  pages = {020356},
  numpages = {12},
  year = {2025},
  month = {Jun},
  publisher = {American Physical Society},
  doi = {10.1103/thxx-njr6},
  url = {https://link.aps.org/doi/10.1103/thxx-njr6}
}

@misc{GSJ24,
      title={Magic state cultivation: growing {T} states as cheap as {CNOT} gates}, 
      author={Craig Gidney and Noah Shutty and Cody Jones},
      year={2024},
      eprint={2409.17595},
      archivePrefix={arXiv},
      primaryClass={quant-ph},
      url={https://arxiv.org/abs/2409.17595}, 
}

@article{STC+26,
  title = {Fold-transversal surface code cultivation},
  author = {Sahay, Kaavya and Tsai, Pei-Kai and Chang, Kathleen (Katie) and Su, Qile and Smith, Thomas B. and Singh, Shraddha and Puri, Shruti},
  journal = {PRX Quantum},
  volume = {7},
  issue = {3},
  pages = {033006},
  numpages = {26},
  year = {2026},
  month = {Jul},
  publisher = {American Physical Society},
  doi = {10.1103/gpvl-lg4c},
  url = {https://link.aps.org/doi/10.1103/gpvl-lg4c}
}

@misc{HBW26,
      title={Constant depth magic state cultivation with {C}lifford measurements by gauging}, 
      author={Bence Hetényi and Benjamin J. Brown and Dominic J. Williamson},
      year={2026},
      eprint={2603.05429},
      archivePrefix={arXiv},
      primaryClass={quant-ph},
      url={https://arxiv.org/abs/2603.05429}, 
}

@article{VJG+26,
  title = {High Rate Magic State Cultivation on the Surface Code},
  author = {Vaknin, Yotam and Jacoby, Shoham and Grimsmo, Arne and Retzker, Alex},
  journal = {PRX Quantum},
  volume = {7},
  issue = {1},
  pages = {010353},
  numpages = {20},
  year = {2026},
  month = {Mar},
  publisher = {American Physical Society},
  doi = {10.1103/p8tw-6kq9},
  url = {https://link.aps.org/doi/10.1103/p8tw-6kq9}
}

@article{CCL+26,
  title = {Efficient Magic State Cultivation on {${\mathbb{R}\mathbb{P}}^{2}$}},
  author = {Chen, Zi-Han and Chen, Ming-Cheng and Lu, Chao-Yang and Pan, Jian-Wei},
  journal = {PRX Quantum},
  volume = {7},
  issue = {1},
  pages = {010315},
  numpages = {19},
  year = {2026},
  month = {Jan},
  publisher = {American Physical Society},
  doi = {10.1103/9kys-3whh},
  url = {https://link.aps.org/doi/10.1103/9kys-3whh}
}

@misc{HTI+25,
      title={Efficient magic state cultivation with lattice surgery}, 
      author={Yutaka Hirano and Riki Toshio and Tomohiro Itogawa and Keisuke Fujii},
      year={2025},
      eprint={2510.24615},
      archivePrefix={arXiv},
      primaryClass={quant-ph},
      url={https://arxiv.org/abs/2510.24615}, 
}

@misc{CFS26,
      title={Efficient Magic State Cultivation for $\sqrt{T}$ Gates}, 
      author={I-Chi Chen and Matheus da Silva Fonseca and Andrew Sornborger},
      year={2026},
      eprint={2606.10430},
      archivePrefix={arXiv},
      primaryClass={quant-ph},
      url={https://arxiv.org/abs/2606.10430}, 
}

@article{WZZ26,
  doi = {10.22331/q-2026-06-12-2134},
  url = {https://doi.org/10.22331/q-2026-06-12-2134},
  title = {Simulating magic state cultivation with few {C}lifford terms},
  author = {Wan, Kwok Ho and Zhong, Zhenghao and Zapirain, Ainhoa},
  journal = {{Quantum}},
  issn = {2521-327X},
  publisher = {{Verein zur F{\"{o}}rderung des Open Access Publizierens in den Quantenwissenschaften}},
  volume = {10},
  pages = {2134},
  month = jun,
  year = {2026}
}

@article{SDK26,
  title = {Efficient Simulation of Logical Magic State Preparation Protocols},
  author = {Surti, Samyak and Daguerre, Lucas and Kim, Isaac H.},
  journal = {PRX Quantum},
  volume = {7},
  issue = {2},
  pages = {020329},
  numpages = {32},
  year = {2026},
  month = {May},
  publisher = {American Physical Society},
  doi = {10.1103/fby6-xjbm},
  url = {https://link.aps.org/doi/10.1103/fby6-xjbm}
}

@article{BSH+21,
  doi = {10.22331/q-2021-11-16-580},
  url = {https://doi.org/10.22331/q-2021-11-16-580},
  title = {Clifford {C}ircuit {O}ptimization with {T}emplates and {S}ymbolic {P}auli {G}ates},
  author = {Bravyi, Sergey and Shaydulin, Ruslan and Hu, Shaohan and Maslov, Dmitri},
  journal = {{Quantum}},
  issn = {2521-327X},
  volume = {5},
  pages = {580},
  month = nov,
  year = {2021}
}

@misc{ZC19,
      title={Optimizing {T} gates in {C}lifford+{T} circuit as $\pi/4$ rotations around Paulis}, 
      author={Fang Zhang and Jianxin Chen},
      year={2019},
      eprint={1903.12456},
      archivePrefix={arXiv},
      primaryClass={quant-ph},
      url={https://arxiv.org/abs/1903.12456}, 
}

@article{AB06,
  title = {Fast simulation of stabilizer circuits using a graph-state representation},
  author = {Anders, Simon and Briegel, Hans J.},
  journal = {Phys. Rev. A},
  volume = {73},
  issue = {2},
  pages = {022334},
  numpages = {9},
  year = {2006},
  month = {Feb},
  publisher = {American Physical Society},
  doi = {10.1103/PhysRevA.73.022334},
  url = {https://link.aps.org/doi/10.1103/PhysRevA.73.022334}
}

@article{AG04,
  title = {Improved simulation of stabilizer circuits},
  author = {Aaronson, Scott and Gottesman, Daniel},
  journal = {Phys. Rev. A},
  volume = {70},
  issue = {5},
  pages = {052328},
  numpages = {14},
  year = {2004},
  month = {Nov},
  publisher = {American Physical Society},
  doi = {10.1103/PhysRevA.70.052328},
  url = {https://link.aps.org/doi/10.1103/PhysRevA.70.052328}
}

@article{BH22,
  doi = {10.22331/q-2022-09-15-803},
  url = {https://doi.org/10.22331/q-2022-09-15-803},
  title = {Fast {S}tabiliser {S}imulation with {Q}uadratic {F}orm {E}xpansions},
  author = {Beaudrap, Niel de and Herbert, Steven},
  journal = {{Quantum}},
  issn = {2521-327X},
  publisher = {{Verein zur F{\"{o}}rderung des Open Access Publizierens in den Quantenwissenschaften}},
  volume = {6},
  pages = {803},
  month = sep,
  year = {2022}
}

@misc{DP23,
      title={Simulation of noisy {C}lifford circuits without fault propagation}, 
      author={Nicolas Delfosse and Adam Paetznick},
      year={2023},
      eprint={2309.15345},
      archivePrefix={arXiv},
      primaryClass={quant-ph},
      url={https://arxiv.org/abs/2309.15345}, 
}

@misc{GLW+25,
      title={{STABSim}: A Parallelized {C}lifford Simulator with Features Beyond Direct Simulation}, 
      author={Sean Garner and Chenxu Liu and Meng Wang and Samuel Stein and Ang Li},
      year={2025},
      eprint={2507.03092},
      archivePrefix={arXiv},
      primaryClass={quant-ph},
      url={https://arxiv.org/abs/2507.03092}, 
}

@misc{OTL26,
      title={{GPU}-Accelerated Quantum Simulation of Stabilizer Circuits}, 
      author={Muhammad Osama and Dimitrios Thanos and Alfons Laarman},
      year={2026},
      eprint={2603.14641},
      archivePrefix={arXiv},
      primaryClass={quant-ph},
      url={https://arxiv.org/abs/2603.14641}, 
}

@ARTICLE{GM15,
  author={García, Héctor J. and Markov, Igor L.},
  journal={IEEE Transactions on Computers}, 
  title={Simulation of Quantum Circuits via Stabilizer Frames}, 
  year={2015},
  volume={64},
  number={8},
  pages={2323--2336},
  doi={10.1109/TC.2014.2360532}}

@misc{GMC13,
      title={Efficient Inner-product Algorithm for Stabilizer States}, 
      author={Hector J. Garcia and Igor L. Markov and Andrew W. Cross},
      year={2013},
      eprint={1210.6646},
      archivePrefix={arXiv},
      primaryClass={cs.ET},
      url={https://arxiv.org/abs/1210.6646}, 
}

@misc{Yoder12,
  title={A generalization of the stabilizer formalism for simulating arbitrary quantum circuits},
  author={Yoder, Theodore J},
  url={http://www.scottaaronson.com/showcase2/report/ted-yoder.pdf},
  year={2012}
}

@article{DD24,
doi = {10.1088/1751-8121/ad8607},
url = {https://doi.org/10.1088/1751-8121/ad8607},
year = {2024},
month = {oct},
publisher = {IOP Publishing},
volume = {57},
number = {45},
pages = {455301},
author = {Descamps, Éloi and Dakić, Borivoje},
title = {On the stabilizer formalism and its generalization},
journal = {Journal of Physics A: Mathematical and Theoretical},
}

@misc{HDD+26,
      title={Qimax: Efficient quantum simulation via {GPU}-accelerated extended stabilizer formalism}, 
      author={Vu Tuan Hai and Bui Cao Doanh and Le Vu Trung Duong and Pham Hoai Luan and Yasuhiko Nakashima},
      year={2026},
      eprint={2505.03307},
      archivePrefix={arXiv},
      primaryClass={quant-ph},
      url={https://arxiv.org/abs/2505.03307}, 
}

@article{BBC+19,
  doi = {10.22331/q-2019-09-02-181},
  url = {https://doi.org/10.22331/q-2019-09-02-181},
  title = {Simulation of quantum circuits by low-rank stabilizer decompositions},
  author = {Bravyi, Sergey and Browne, Dan and Calpin, Padraic and Campbell, Earl and Gosset, David and Howard, Mark},
  journal = {{Quantum}},
  issn = {2521-327X},
  publisher = {{Verein zur F{\"{o}}rderung des Open Access Publizierens in den Quantenwissenschaften}},
  volume = {3},
  pages = {181},
  month = sep,
  year = {2019}
}

@inproceedings{MT24,
author = {Mehraban, Saeed and Tahmasbi, Mehrdad},
title = {Quadratic Lower Bounds on the Approximate Stabilizer Rank: A Probabilistic Approach},
year = {2024},
isbn = {9798400703836},
url = {https://doi.org/10.1145/3618260.3649733},
doi = {10.1145/3618260.3649733},
booktitle = {Proceedings of the 56th Annual ACM Symposium on Theory of Computing},
pages = {608--619},
numpages = {12},
}

@article{BG16,
  title = {Improved Classical Simulation of Quantum Circuits Dominated by Clifford Gates},
  author = {Bravyi, Sergey and Gosset, David},
  journal = {Phys. Rev. Lett.},
  volume = {116},
  issue = {25},
  pages = {250501},
  numpages = {5},
  year = {2016},
  month = {Jun},
  publisher = {American Physical Society},
  doi = {10.1103/PhysRevLett.116.250501},
  url = {https://link.aps.org/doi/10.1103/PhysRevLett.116.250501}
}

@article{BSS16,
  title = {Trading Classical and Quantum Computational Resources},
  author = {Bravyi, Sergey and Smith, Graeme and Smolin, John A.},
  journal = {Phys. Rev. X},
  volume = {6},
  issue = {2},
  pages = {021043},
  numpages = {14},
  year = {2016},
  month = {Jun},
  publisher = {American Physical Society},
  doi = {10.1103/PhysRevX.6.021043},
  url = {https://link.aps.org/doi/10.1103/PhysRevX.6.021043}
}

@article{HL19,
  title = {Approximate stabilizer rank and improved weak simulation of Clifford-dominated circuits for qudits},
  author = {Huang, Yifei and Love, Peter},
  journal = {Phys. Rev. A},
  volume = {99},
  issue = {5},
  pages = {052307},
  numpages = {13},
  year = {2019},
  month = {May},
  publisher = {American Physical Society},
  doi = {10.1103/PhysRevA.99.052307},
  url = {https://link.aps.org/doi/10.1103/PhysRevA.99.052307}
}

@article{QPG21,
  doi = {10.22331/q-2021-12-20-606},
  url = {https://doi.org/10.22331/q-2021-12-20-606},
  title = {Improved upper bounds on the stabilizer rank of magic states},
  author = {Qassim, Hammam and Pashayan, Hakop and Gosset, David},
  journal = {{Quantum}},
  issn = {2521-327X},
  publisher = {{Verein zur F{\"{o}}rderung des Open Access Publizierens in den Quantenwissenschaften}},
  volume = {5},
  pages = {606},
  month = dec,
  year = {2021}
}

@article{SRP+21,
  title = {Quantifying Quantum Speedups: Improved Classical Simulation From Tighter Magic Monotones},
  author = {Seddon, James R. and Regula, Bartosz and Pashayan, Hakop and Ouyang, Yingkai and Campbell, Earl T.},
  journal = {PRX Quantum},
  volume = {2},
  issue = {1},
  pages = {010345},
  numpages = {42},
  year = {2021},
  month = {Mar},
  publisher = {American Physical Society},
  doi = {10.1103/PRXQuantum.2.010345},
  url = {https://link.aps.org/doi/10.1103/PRXQuantum.2.010345}
}

@article{PRK22,
  title = {Fast Estimation of Outcome Probabilities for Quantum Circuits},
  author = {Pashayan, Hakop and Reardon-Smith, Oliver and Korzekwa, Kamil and Bartlett, Stephen D.},
  journal = {PRX Quantum},
  volume = {3},
  issue = {2},
  pages = {020361},
  numpages = {29},
  year = {2022},
  month = {Jun},
  publisher = {American Physical Society},
  doi = {10.1103/PRXQuantum.3.020361},
  url = {https://link.aps.org/doi/10.1103/PRXQuantum.3.020361}
}

@misc{Kocia22,
      title={Improved Strong Simulation of Universal Quantum Circuits}, 
      author={Lucas Kocia},
      year={2022},
      eprint={2012.11739},
      archivePrefix={arXiv},
      primaryClass={quant-ph},
      url={https://arxiv.org/abs/2012.11739}, 
}

@InProceedings{KWV22,
  author =	{Kissinger, Aleks and van de Wetering, John and Vilmart, Renaud},
  title =	{{Classical Simulation of Quantum Circuits with Partial and Graphical Stabiliser Decompositions}},
  booktitle =	{17th Conference on the Theory of Quantum Computation, Communication and Cryptography (TQC 2022)},
  pages =	{5:1--5:13},
  ISBN =	{978-3-95977-237-2},
  ISSN =	{1868-8969},
  year =	{2022},
  volume =	{232},
  URL =		{https://drops.dagstuhl.de/entities/document/10.4230/LIPIcs.TQC.2022.5},
  URN =		{urn:nbn:de:0030-drops-165128},
  doi =		{10.4230/LIPIcs.TQC.2022.5},
}

@article{KW22,
doi = {10.1088/2058-9565/ac5d20},
url = {https://doi.org/10.1088/2058-9565/ac5d20},
year = {2022},
month = {jul},
publisher = {IOP Publishing},
volume = {7},
number = {4},
pages = {044001},
author = {Kissinger, Aleks and van de Wetering, John},
title = {Simulating quantum circuits with {ZX}-calculus reduced stabiliser decompositions},
journal = {Quantum Science and Technology},
}

@article{SK25,
   title={Fast Classical Simulation of Quantum Circuits via Parametric Rewriting in the ZX-Calculus},
   volume={426},
   ISSN={2075-2180},
   url={http://dx.doi.org/10.4204/EPTCS.426.10},
   DOI={10.4204/eptcs.426.10},
   journal={Electronic Proceedings in Theoretical Computer Science},
   publisher={Open Publishing Association},
   author={Sutcliffe, Matthew and Kissinger, Aleks},
   year={2025},
   month=Aug, pages={247--269} }

@article{SK24,
   title={Procedurally Optimised {ZX}-Diagram Cutting for Efficient {T}-Decomposition in Classical Simulation},
   volume={406},
   ISSN={2075-2180},
   url={http://dx.doi.org/10.4204/EPTCS.406.3},
   DOI={10.4204/eptcs.406.3},
   journal={Electronic Proceedings in Theoretical Computer Science},
   publisher={Open Publishing Association},
   author={Sutcliffe, Matthew and Kissinger, Aleks},
   year={2024},
   month=Aug, pages={63--78} }

@article{MG24,
  title = {Stabilizer Tensor Networks: Universal Quantum Simulator on a Basis of Stabilizer States},
  author = {Masot-Llima, Sergi and Garcia-Saez, Artur},
  journal = {Phys. Rev. Lett.},
  volume = {133},
  issue = {23},
  pages = {230601},
  numpages = {6},
  year = {2024},
  month = {Dec},
  publisher = {American Physical Society},
  doi = {10.1103/PhysRevLett.133.230601},
  url = {https://link.aps.org/doi/10.1103/PhysRevLett.133.230601}
}

@article{NHW25,
  title = {Stabilizer Tensor Networks with Magic State Injection},
  author = {Nakhl, Azar C. and Harper, Ben and West, Maxwell and Dowling, Neil and Sevior, Martin and Quella, Thomas and Usman, Muhammad},
  journal = {Phys. Rev. Lett.},
  volume = {134},
  issue = {19},
  pages = {190602},
  numpages = {9},
  year = {2025},
  month = {May},
  publisher = {American Physical Society},
  doi = {10.1103/PhysRevLett.134.190602},
  url = {https://link.aps.org/doi/10.1103/PhysRevLett.134.190602}
}

@misc{UA26,
      title={{SyQMA}: A memory-efficient, symbolic and exact universal simulator for quantum error correction}, 
      author={George Umbrarescu and David Amaro},
      year={2026},
      eprint={2604.15043},
      archivePrefix={arXiv},
      primaryClass={quant-ph},
      url={https://arxiv.org/abs/2604.15043}, 
}

@article{RLC19,
  title = {Simulation of qubit quantum circuits via Pauli propagation},
  author = {Rall, Patrick and Liang, Daniel and Cook, Jeremy and Kretschmer, William},
  journal = {Phys. Rev. A},
  volume = {99},
  issue = {6},
  pages = {062337},
  numpages = {10},
  year = {2019},
  month = {Jun},
  publisher = {American Physical Society},
  doi = {10.1103/PhysRevA.99.062337},
  url = {https://link.aps.org/doi/10.1103/PhysRevA.99.062337}
}

@misc{RJT+26,
      title={Pauli Propagation: A Computational Framework for Simulating Quantum Systems}, 
      author={Manuel S. Rudolph and Tyson Jones and Yanting Teng and Armando Angrisani and Zoë Holmes},
      year={2026},
      eprint={2505.21606},
      archivePrefix={arXiv},
      primaryClass={quant-ph},
      url={https://arxiv.org/abs/2505.21606}, 
}

@misc{HOR+26,
      title={Simulating Quantum Error Correction beyond Pauli Stochastic Errors}, 
      author={Jordan Hines and Corey Ostrove and Kenneth Rudinger and Stefan Seritan and Kevin Young and Robin Blume-Kohout and Timothy Proctor},
      year={2026},
      eprint={2603.18457},
      archivePrefix={arXiv},
      primaryClass={quant-ph},
      url={https://arxiv.org/abs/2603.18457}, 
}

@article{AMR26,
  title = {Simulating Quantum Circuits with Arbitrary Local Noise Using Pauli Propagation},
  author = {Angrisani, Armando and Mele, Antonio A. and Rudolph, Manuel S. and Cerezo, M. and Holmes, Zo\"e},
  journal = {PRX Quantum},
  volume = {7},
  issue = {2},
  pages = {020313},
  numpages = {46},
  year = {2026},
  month = {Apr},
  publisher = {American Physical Society},
  doi = {10.1103/fb28-wlv2},
  url = {https://link.aps.org/doi/10.1103/fb28-wlv2}
}

@article{DTB+25,
  doi = {10.22331/q-2025-11-06-1905},
  url = {https://doi.org/10.22331/q-2025-11-06-1905},
  title = {Designing fault-tolerant circuits using detector error models},
  author = {Derks, Peter-Jan H.S. and Townsend-Teague, Alex and Burchards, Ansgar G. and Eisert, Jens},
  journal = {{Quantum}},
  issn = {2521-327X},
  publisher = {{Verein zur F{\"{o}}rderung des Open Access Publizierens in den Quantenwissenschaften}},
  volume = {9},
  pages = {1905},
  month = nov,
  year = {2025}
}

@article{BV13,
  title = {Simulation of rare events in quantum error correction},
  author = {Bravyi, Sergey and Vargo, Alexander},
  journal = {Phys. Rev. A},
  volume = {88},
  issue = {6},
  pages = {062308},
  numpages = {14},
  year = {2013},
  month = {Dec},
  publisher = {American Physical Society},
  doi = {10.1103/PhysRevA.88.062308},
  url = {https://link.aps.org/doi/10.1103/PhysRevA.88.062308}
}

@article{HWR24,
  title = {Dynamical subset sampling of quantum error-correcting protocols},
  author = {Heu\ss{}en, Sascha and Winter, Don and Rispler, Manuel and M\"uller, Markus},
  journal = {Phys. Rev. Res.},
  volume = {6},
  issue = {1},
  pages = {013177},
  numpages = {26},
  year = {2024},
  month = {Feb},
  publisher = {American Physical Society},
  doi = {10.1103/PhysRevResearch.6.013177},
  url = {https://link.aps.org/doi/10.1103/PhysRevResearch.6.013177}
}

@misc{MGO25,
      title={Rare Event Simulation of Quantum Error-Correcting Circuits}, 
      author={Carolyn Mayer and Anand Ganti and Uzoma Onunkwo and Tzvetan Metodi and Benjamin Anker and Jacek Skryzalin},
      year={2025},
      eprint={2509.13678},
      archivePrefix={arXiv},
      primaryClass={quant-ph},
      url={https://arxiv.org/abs/2509.13678}, 
}

@misc{BCC+25,
      title={Fail fast: techniques to probe rare events in quantum error correction}, 
      author={Michael E. Beverland and Malcolm Carroll and Andrew W. Cross and Theodore J. Yoder},
      year={2025},
      eprint={2511.15177},
      archivePrefix={arXiv},
      primaryClass={quant-ph},
      url={https://arxiv.org/abs/2511.15177}, 
}

@misc{YP26,
      title={Scalable testing of quantum error correction}, 
      author={John Zhuoyang Ye and Jens Palsberg},
      year={2026},
      eprint={2602.04921},
      archivePrefix={arXiv},
      primaryClass={quant-ph},
      url={https://arxiv.org/abs/2602.04921}, 
}

@misc{MOH+25,
      title={Efficient simulation of Clifford circuits with small Markovian errors}, 
      author={Ashe Miller and Corey Ostrove and Jordan Hines and Robin Blume-Kohout and Kevin Young and Timothy Proctor},
      year={2025},
      eprint={2504.15128},
      archivePrefix={arXiv},
      primaryClass={quant-ph},
      url={https://arxiv.org/abs/2504.15128}, 
}

@article{ADP14,
  title = {Fault-Tolerant Conversion between the {S}teane and {R}eed-{M}uller Quantum Codes},
  author = {Anderson, Jonas T. and Duclos-Cianci, Guillaume and Poulin, David},
  journal = {Phys. Rev. Lett.},
  volume = {113},
  issue = {8},
  pages = {080501},
  numpages = {5},
  year = {2014},
  month = {Aug},
  publisher = {American Physical Society},
  doi = {10.1103/PhysRevLett.113.080501},
  url = {https://link.aps.org/doi/10.1103/PhysRevLett.113.080501}
}

@article{Hector15,
  title={Gauge color codes: optimal transversal gates and gauge fixing in topological stabilizer codes},
  author={Bomb{\'\i}n, H{\'e}ctor},
  journal={New Journal of Physics},
  volume={17},
  number={8},
  pages={083002},
  year={2015},
  publisher={IOP Publishing},
  doi={10.1088/1367-2630/17/8/083002}
}

@article{KB15,
  title = {Universal transversal gates with color codes: A simplified approach},
  author = {Kubica, Aleksander and Beverland, Michael E.},
  journal = {Phys. Rev. A},
  volume = {91},
  issue = {3},
  pages = {032330},
  numpages = {12},
  year = {2015},
  month = {Mar},
  publisher = {American Physical Society},
  doi = {10.1103/PhysRevA.91.032330},
  url = {https://link.aps.org/doi/10.1103/PhysRevA.91.032330}
}

@article{PR13,
  title = {Universal Fault-Tolerant Quantum Computation with Only Transversal Gates and Error Correction},
  author = {Paetznick, Adam and Reichardt, Ben W.},
  journal = {Phys. Rev. Lett.},
  volume = {111},
  issue = {9},
  pages = {090505},
  numpages = {5},
  year = {2013},
  month = {Aug},
  publisher = {American Physical Society},
  doi = {10.1103/PhysRevLett.111.090505},
  url = {https://link.aps.org/doi/10.1103/PhysRevLett.111.090505}
}

@article{JL14,
  title = {Using Concatenated Quantum Codes for Universal Fault-Tolerant Quantum Gates},
  author = {Jochym-O'Connor, Tomas and Laflamme, Raymond},
  journal = {Phys. Rev. Lett.},
  volume = {112},
  issue = {1},
  pages = {010505},
  numpages = {5},
  year = {2014},
  month = {Jan},
  publisher = {American Physical Society},
  doi = {10.1103/PhysRevLett.112.010505},
  url = {https://link.aps.org/doi/10.1103/PhysRevLett.112.010505}
}

@article{CJL16,
  title = {Thresholds for Universal Concatenated Quantum Codes},
  author = {Chamberland, Christopher and Jochym-O'Connor, Tomas and Laflamme, Raymond},
  journal = {Phys. Rev. Lett.},
  volume = {117},
  issue = {1},
  pages = {010501},
  numpages = {5},
  year = {2016},
  month = {Jun},
  publisher = {American Physical Society},
  doi = {10.1103/PhysRevLett.117.010501},
  url = {https://link.aps.org/doi/10.1103/PhysRevLett.117.010501}
}

@article{TYC16,
  title = {Universal Fault-Tolerant Gates on Concatenated Stabilizer Codes},
  author = {Yoder, Theodore J. and Takagi, Ryuji and Chuang, Isaac L.},
  journal = {Phys. Rev. X},
  volume = {6},
  issue = {3},
  pages = {031039},
  numpages = {23},
  year = {2016},
  month = {Sep},
  publisher = {American Physical Society},
  doi = {10.1103/PhysRevX.6.031039},
  url = {https://link.aps.org/doi/10.1103/PhysRevX.6.031039}
}

@article{TYC17,
  title = {Error rates and resource overheads of encoded three-qubit gates},
  author = {Takagi, Ryuji and Yoder, Theodore J. and Chuang, Isaac L.},
  journal = {Phys. Rev. A},
  volume = {96},
  issue = {4},
  pages = {042302},
  numpages = {13},
  year = {2017},
  month = {Oct},
  publisher = {American Physical Society},
  doi = {10.1103/PhysRevA.96.042302},
  url = {https://link.aps.org/doi/10.1103/PhysRevA.96.042302}
}

@misc{AMM26,
      title={Analytical and Compressed Simulation of Noisy Stabilizer Circuits}, 
      author={Paul Aigner and Jasmin Matti and Maria Flors Mor-Ruiz and Julius Wallnöfer and Wolfgang Dür},
      year={2026},
      eprint={2604.22588},
      archivePrefix={arXiv},
      primaryClass={quant-ph},
      url={https://arxiv.org/abs/2604.22588}, 
}

@misc{CRG+26,
      title={{MPStab}: an hybrid stabilizers tensor-network quantum circuit simulator}, 
      author={Giulio Crognaletti and Mattia Robbiano and Michele Grossi and Matteo Robbiati},
      year={2026},
      eprint={2607.24258},
      archivePrefix={arXiv},
      primaryClass={quant-ph},
      url={https://arxiv.org/abs/2607.24258}, 
}

@misc{MBK26,
author = "M{\"u}ller, Leon and B{\"a}rligea, Adelina and Knapp, Alexander and Kottmann, Jakob S.",
title = "PauliEngine: High-Performant Symbolic Arithmetic for Quantum Operations",
year = 2026,
doi = "10.18420/se2026-ws_20",
howpublished = "SE2026",
publisher = "Gesellschaft für Informatik, Bonn",
}

@misc{K26,
    title = {{PauLIB}: A High-Performance Library for Processing Pauli Strings},
    author = {Kr{\"o}tz, Florian},
    year={2026},
    eprint={2605.25974},
    archivePrefix={arXiv},
    primaryClass={quant-ph},
    url={https://arxiv.org/abs/2605.25974}, 
}

@inproceedings{AGL+23,
author = {Aharonov, Dorit and Gao, Xun and Landau, Zeph and Liu, Yunchao and Vazirani, Umesh},
title = {A Polynomial-Time Classical Algorithm for Noisy Random Circuit Sampling},
year = {2023},
isbn = {9781450399135},
url = {https://doi.org/10.1145/3564246.3585234},
doi = {10.1145/3564246.3585234},
booktitle = {Proceedings of the 55th Annual ACM Symposium on Theory of Computing},
pages = {945--957},
numpages = {13},
series = {STOC 2023}
}

@article{FRD+25,
  title={Classical simulations of noisy variational quantum circuits},
  author={Fontana, Enrico and Rudolph, Manuel S and Duncan, Ross and Rungger, Ivan and C{\^\i}rstoiu, Cristina},
  journal={npj Quantum Information},
  volume={11},
  number={1},
  pages={84},
  year={2025},
  publisher={Nature Publishing Group UK London},
  doi={10.1038/s41534-024-00955-1}
}

@article{SWC+24,
  title = {Simulating Noisy Variational Quantum Algorithms: A Polynomial Approach},
  author = {Shao, Yuguo and Wei, Fuchuan and Cheng, Song and Liu, Zhengwei},
  journal = {Phys. Rev. Lett.},
  volume = {133},
  issue = {12},
  pages = {120603},
  numpages = {7},
  year = {2024},
  month = {Sep},
  publisher = {American Physical Society},
  doi = {10.1103/PhysRevLett.133.120603},
  url = {https://link.aps.org/doi/10.1103/PhysRevLett.133.120603}
}

@article{SYG+25,
  title = {A Polynomial-Time Classical Algorithm for Noisy Quantum Circuits},
  author = {Schuster, Thomas and Yin, Chao and Gao, Xun and Yao, Norman Y.},
  journal = {Phys. Rev. X},
  volume = {15},
  issue = {4},
  pages = {041018},
  numpages = {31},
  year = {2025},
  month = {Nov},
  publisher = {American Physical Society},
  doi = {10.1103/xct1-7kf2},
  url = {https://link.aps.org/doi/10.1103/xct1-7kf2}
}

@article{GCT25,
  doi = {10.22331/q-2025-05-05-1730},
  url = {https://doi.org/10.22331/q-2025-05-05-1730},
  title = {Pauli path simulations of noisy quantum circuits beyond average case},
  author = {Gonz{\'{a}}lez-Garc{\'{i}}a, Guillermo and Cirac, J. Ignacio and Trivedi, Rahul},
  journal = {{Quantum}},
  issn = {2521-327X},
  publisher = {{Verein zur F{\"{o}}rderung des Open Access Publizierens in den Quantenwissenschaften}},
  volume = {9},
  pages = {1730},
  month = may,
  year = {2025}
}

\end{document}